\documentclass[11pt]{article}
\pdfoutput=1

\usepackage{jheppub}
\usepackage{url,comment}
\usepackage{times}
\usepackage{latexsym}
\usepackage{graphicx, graphics, hyperref, amsmath, amssymb, slashed, color, bbm,amsthm,multirow,caption,subcaption}
 
\newcommand{\nc}{\newcommand}

\nc{\beq}{\begin{equation}}
\nc{\eeq}{\end{equation}}
\nc{\barray}{\begin{eqnarray}}
\nc{\earray}{\end{eqnarray}}
\nc{\barrayn}{\begin{eqnarray*}}
\nc{\earrayn}{\end{eqnarray*}}
\nc{\bcenter}{\begin{center}}
\nc{\ecenter}{\end{center}}
\nc{\mc}{\mathcal}
\nc{\er}[1]{(\ref{eq:#1})}
\nc{\onehalf}{\frac{1}{2}}
\nc{\partialbar}{\bar{\partial}}
\nc{\psit}{\widetilde{\psi}}
\nc{\Tr}{\mbox{Tr}}
\nc{\hc}{\mbox{H.c.}}
\nc{\ev}{\;\mathrm{eV}}
\nc{\mev}{\;\mathrm{MeV}}
\nc{\gev}{\;\mathrm{GeV}}
\nc{\tev}{\;\mathrm{TeV}}

\def\chii0{\chi_i^0}
\def\chij0{\chi_j^0}

\newcommand{\gsim}{\lower.7ex\hbox{$\;\stackrel{\textstyle>}{\sim}\;$}}
\newcommand{\lsim}{\lower.7ex\hbox{$\;\stackrel{\textstyle<}{\sim}\;$}}
\nc{\ttbar}{t\bar t}

\nc{\Lag}{\mathcal{L}}

\nc{\vev}[1]{\left<#1\right>}
\nc{\abs}[1]{\left|#1\right|}

\nc{\mpl}{M_{Pl}}
\nc{\sv}{\langle \sigma v\rangle}
\nc{\too}{\leftrightarrow}
\newcommand{\lp}{\left(}
\newcommand{\rp}{\right)} 
\nc{\zd}{{Z_D}}
\nc{\order}[1]{\mathcal O \left(#1 \right)}

\newcommand{\twobytwomatrix}[4]
{
	\left(
		\begin{array}{cc}
			#1 &  #2 \\
			#3 &  #4
		\end{array}
	\right)
}

\title{Leak-in Dark Matter}

\author[a]{Jared A.~Evans,}
\author[b]{Cristian Gaidau,}
\author[b]{and Jessie Shelton}
\affiliation[a]{Department of Physics, University of Cincinnati, Cincinnati, Ohio 45221, USA}
\affiliation[b]{Department of Physics, University of Illinois at Urbana-Champaign, Urbana, IL 61801, USA}

\emailAdd{jaredaevans@gmail.com}
\emailAdd{gaidau2@illinois.edu}
\emailAdd{sheltonj@illinois.edu}

\abstract{
We introduce leak-in dark matter, a novel out-of-equilibrium origin for the dark matter (DM) in the universe.   We provide a comprehensive and unified discussion of a minimal, internally-thermalized, hidden sector populated from an out-of-equilibrium, feeble connection to the hotter standard model (SM) sector.  We emphasize that when this out-of-equilibrium interaction is renormalizable, the colder sector undergoes an extended phase of non-adiabatic evolution largely independent of initial conditions, which we dub ``leak-in.''   We discuss the leak-in phase in generality, and establish the general properties of dark matter that freezes out from a radiation bath undergoing such a leak-in phase.  As a concrete example, we consider a model where the SM has an out-of-equilibrium $B-L$ vector portal interaction with a minimal hidden sector.  We discuss the interplay between leak-in and freezein processes in this theory in detail and demonstrate regions where leak-in yields the full relic abundance.  We study observational prospects for $B-L$ vector portal leak-in DM, and find that despite the requisite small coupling to the SM, a variety of experiments can serve as sensitive probes of leak-in dark matter.   Additionally, regions allowed by all current constraints yield DM with self-interactions large enough to address small-scale structure anomalies. 
}

\arxivnumber{nnnn.nnnnn}

\begin{document}

\maketitle

\section{Introduction}\label{sec:intro}

Despite overwhelming gravitational evidence for the existence of dark matter (DM), the particle properties of DM remain mysterious.   Historically, one of the best-motivated candidates for particle DM has been a weakly-interacting massive particle (WIMP), or more generally, a particle that was in thermal equilibrium with the standard model (SM) plasma in the early universe, but froze out as number-changing interactions with the SM, e.g.~annihilations DM DM $\to$ SM SM, departed from equilibrium.  One major appealing feature of this class of models is that the DM relic abundance is directly tied to its couplings to the SM, giving rise to definite and accessible experimental targets.  Owing to the spectacular success of experiments searching for DM---in direct, collider, and indirect probes---the surviving WIMP parameter space is rapidly shrinking.  Other scenarios for the origin of particle dark matter, and their resulting experimental signatures, are thus of high interest.

One broad and generic scenario for the origin of  DM is that its relic abundance can be determined by interactions within an internally thermalized hidden sector (HS), with minimal direct involvement of SM fields \cite{Kolb:1985bf,Hodges:1993yb,Pospelov:2007mp, Feng:2008ya, Feng:2008mu,ArkaniHamed:2008qn, Shelton:2010ta, Haba:2010bm,Buckley:2010ui}.
Such self-interacting hidden sectors open many avenues for addressing long-standing mysteries in both particle and astrophysics, and can predict qualitatively novel signatures.  More broadly, internally thermalized hidden sectors are a simple and generic possible source for the DM of our universe, and it is worth addressing in some generality how the possible cosmological origin stories for such hidden sectors impact the dynamics and signatures of the DM they produce.

One minimal and predictive way to populate a thermal dark radiation bath in the early universe is by producing it directly from interactions with the SM radiation bath.  In the simplest scenarios, 
these interactions are sufficiently strong to bring the hidden sector into thermal equilibrium with the SM.   In this paper we will focus on the regime where the leading interaction between the two sectors never enters equilibrium.   

In this scenario, feeble interactions allow energy to leak from the hot SM radiation bath into the colder hidden sector.  When the leading interaction is non-renormalizable, the energy injection from the SM rapidly becomes negligible as the universe expands. The population of the hidden sector is thus dominated by a limited span of UV temperatures, after which the hidden sector evolves adiabatically \cite{Faraggi:2000pv}.   By contrast, when the leading interaction is renormalizable, energy injection from the SM becomes more and more important as the universe cools.
In this latter case, the hidden sector radiation bath undergoes an extended phase of non-adiabatic evolution that we dub ``leak-in,'' which realizes a quasi-static equilibrium between the energy injection from the SM and the dilution from the expansion of the universe.  The aim of this paper is to investigate the properties of DM that freezes out of a hidden radiation bath in this quasi-static leak-in phase, which we dub ``leak-in dark matter'' (LIDM).  This scenario is distinct from freezein DM \cite{McDonald:2001vt, Hall:2009bx}, where the DM itself is the hidden particle produced from the SM.   The primary difference for leak-in is that the hidden sector is internally thermalized and acquires its own temperature, which fixes the abundances of particles, including dark matter, within the hidden sector. 

Despite the feeble coupling to the SM, there are many potential experimental handles on leak-in dark matter.  
In particular, while LIDM annihilation cross-sections are typically suppressed relative to standard WIMP benchmarks, indirect detection signals are still within reach of a variety of cosmic ray experiments, 
 such as Fermi,  AMS-02,  H.E.S.S., HAWC, CTA, and others \cite{Atwood:2009ez,Aguilar:2013qda,Bernloehr:2003vd,Harding:2015bua,Consortium:2010bc}.   Additionally, observations of the cosmic microwave background (CMB)  can place stringent constraints on LIDM annihilations during recombination.  Direct detection can also be a promising avenue for detecting LIDM, with complementary sensitivity to indirect detection, and XENON1T \cite{Aprile:2018dbl} is currently probing the edge of LIDM parameter space in the benchmark model we will consider later in this work.   There can also be meaningful constraints on the mediator itself, from, for example, stellar cooling or fifth force experiments \cite{Hardy:2016kme}.   Additionally, regions of the LIDM parameter space realize sizable DM self-interaction cross-sections.   Very large DM self-interactions are constrained by dwarf structure and ellipticity, but somewhat smaller self-interactions may be favored by various small-scale structure anomalies \cite{Tulin:2017ara}.

We begin by discussing the general properties of a radiation bath populated by out-of-equilibrium renormalizable interactions in Sec.~\ref{sec:analytics}.  We establish the general properties of DM that freezes out during the resulting leak-in phase in Sec.~\ref{sec:fo}. 
In Sec.~\ref{sec:addingfi}, we introduce a concrete model of a minimal hidden sector, consisting of a feebly-coupled $B-L$ vector boson together with dark matter, and discuss the mechanisms governing the DM relic abundance in detail.  Sec.~\ref{sec:signals} examines the observable signals of the model, with the viable regions of parameter space collected in Sec.~\ref{sec:results}.  We conclude in Sec.~\ref{sec:conclusions}.  Appendices include criteria for attaining internal thermalization in the hidden sector in App.~\ref{sec:inttherm}, some $B-L$ model-building considerations in Sec.~\ref{sec:model}, and details of the energy transfer between SM and hidden sectors in App.~\ref{app:CE}.

\section{Leak-in: the out-of-equilibrium population of a hidden radiation bath}
\label{sec:analytics}

We begin by discussing the out-of-equilibrium population of a dark radiation bath from the SM in some generality.
Throughout this work, we will denote hidden sector (SM) quantities with (without) a tilde.    
The Boltzmann equations describing the temperature evolution of two
internally thermalized radiation baths are 
\begin{eqnarray}
\label{eq:b1}
\dot\rho+4H\rho &= & C_E[\rho, \tilde\rho]\\
\label{eq:b2}
\dot{\tilde\rho}+4H \tilde\rho &= &- C_E[\rho, \tilde\rho]\\
\label{eq:b3}
H^2 & = &\frac {8\pi} {3\mpl^2} (\rho+ \tilde\rho),
\end{eqnarray}
where $C_E$
is the collision term describing the energy transfer 
between sectors, $\rho$ and $\tilde\rho$ are the SM 
and hidden sector energy densities, respectively, and we have made the simplifying approximation of
neglecting the contribution of non-relativistic species to the energy
density.  Assuming that interactions within each sector keep the
sectors in internal thermal equilibrium at separate temperatures $T, \tilde T$,  these
equations can be solved to obtain the dependence of $T$ and $\tilde T$ on the scale factor $a$.

The form of the collision term, and in particular its dependence on $T$ and $\tilde T$, depend on the nature and structure of the leading interaction linking the two sectors. 
When the leading interaction is renormalizable, the collision term falls off more slowly with temperature than the Hubble term: in other words, scattering through renormalizable interactions becomes more important in the late universe relative to the early universe.  This IR-dominance has the useful consequence of making the properties of hidden sectors populated through a renormalizable interaction with the SM relatively insensitive to the unknown reheating temperature of the universe.  

\subsection{
Populating a cold sector through renormalizable interactions}
\label{sec:thermalmix}

There is a limited suite of possible renormalizable operators that allow SM particles to interact with a particle that is a total SM singlet.  These operators include a dark fermion $\psi$ coupling through the neutrino portal, $\mathcal{O}_\nu = \psi H L$,
a dark scalar $S$ coupling through the Higgs portal, $\mathcal{O}_h = \mu S |H|^2+ S^2 |H|^2$, a dark vector boson $Z_D$ coupling through kinetic mixing with hypercharge, $\mathcal{O}_Y = B_{\mu\nu}Z_D^{\mu\nu}$, and a dark vector boson $Z_D$ coupled to the SM through one of the anomaly free currents: either $\mathcal{O}_{L_i-L_j}= Z_{D\mu} J^\mu_{L_i-L_j}$ $(i\neq j)$ or $\mathcal{O}_{B-L}= Z_{D\mu} J^\mu_{B-L}$.
Each of these interactions together with those of the SM generate tree-level $2\to 2$ scattering processes, which are, at zero temperature, independent of the dark particle mass in the $E_{CM}\gg m$ limit.  Dimensional analysis then suggests that the scattering rate in the early universe should scale like $T$, for $T\gg m$, an expectation borne out in explicit kinetic theory calculations \cite{Evans:2017kti}. However, properly accounting for the contribution to the thermal self-energies in the dense medium of the radiation bath can in some cases parametrically alter this expectation \cite{An:2013yfc, Redondo:2013lna, Hardy:2016kme}.
In particular, when a dark species $X$ couples to the SM entirely through mixing with another state in the plasma, there is a parametric suppression of the production rate of $X$ from the SM plasma as $m_X/T \to 0$ \cite{Hardy:2016kme}.  If $X$ can mix with a SM state $A$ in medium, the propagating degrees of freedom can be found by diagonalizing the $2\times 2$ propagation matrix,
\beq
\label{eq:propmatrix}
\twobytwomatrix{\Pi^{AA}}{\Pi^{AX}}{\Pi^{XA}}{m_X^2+\Pi^{XX}},
\eeq
where $\Pi^{IJ}$ are (1PI) thermal self-energies, and for simplicity we have taken $A$ to be massless (in many examples of interest it is the photon).  
The observation of Ref.~\cite{Hardy:2016kme} is that the off-diagonal entries of this matrix provide important corrections to the finite temperature propagator, and therefore to the net production rate of $X$ from the SM plasma.
Taking $X$ to be coupled to the SM plasma through a parameter $\epsilon\ll 1$,  we can write $\Pi^{XX} \equiv \Pi^{XX}_{dk} +\Pi^{XX}_{SM}$, where $\Pi^{XX}_{SM}$ is of order $\mathcal{O}(\epsilon^2)$ and the $\mathcal{O}(\epsilon^0)$ piece of the self-energy, $\Pi^{XX}_{dk}$, accounts for possible contributions from other dark species that may be in the plasma (with no direct coupling to the SM).  Working to leading order in $\epsilon$ and absorbing  $\Pi^{XX}_{dk}$ into an effective mass for the dark state, $\tilde m_X^2$,
 the eigenmodes of Eq.~\ref{eq:propmatrix} are
\begin{eqnarray}
\Pi^{(A)} &=& \Pi^{AA}  +\frac{ (\Pi^{AX})^2}{\Pi^{AA} -\tilde m_X^2} + \mathcal{O}(\epsilon^4)\\
\Pi^{(X)} &=& \tilde m_X^2  +\left(\Pi^{XX}_{SM} - \frac{ (\Pi^{AX})^2}{\Pi^{AA} -\tilde m_X^2} \right) + \mathcal{O}(\epsilon^4).
\label{eq:eigenfreq}
\end{eqnarray}
The production rate of $X$s from the SM plasma is then given in terms of the imaginary part of this self energy,
\beq
\frac{dN_X^{SM}}{dV dt}
   = - \int \frac{d^3 k}{(2\pi)^3}\frac{f_B(E_k)}{E_k}  \mathrm{Im} \left( \Pi^{XX}_{SM} - \frac{ (\Pi^{AX})^2}{\Pi^{AA} -\tilde m^2_X} \right).
\label{eq:prodSM}
\eeq
Now, suppose that $X$ inherits all its couplings to the SM from mixing with $A$. Then we can write (to lowest nontrivial order in $\epsilon$)
\begin{eqnarray}
\nonumber
\Pi^{AA} &=& \mathcal{C}\\
\label{eq:scaling2}
\Pi^{AX} &=& \epsilon \mathcal{C} \\
\nonumber
\Pi^{XX}_{SM} &=& \epsilon^2 \mathcal{C} .
\end{eqnarray}
With this relation, the term in parentheses in Eq.~\ref{eq:prodSM} can be expanded in the $\Pi^{AA}\gg \tilde m_X^2$ limit to obtain
\beq
\label{eq:cancel}
\epsilon^2 \mathcal{C} \left( 1 - 1 \left(1+ \frac{\tilde m_X^2}{\Pi^{AA}}\right) \right) \sim \mathcal{O}(\tilde m_X^2/T^2) \times \epsilon^2 \mathcal{C}
\eeq
which is directly proportional to $\tilde m_X^2$ and vanishes in the $\tilde m/T\to 0$ limit, contrary to the naive expectation from kinetic theory, $\Gamma\propto T$.

On the other hand, if the tight relationship of Eq.~\ref{eq:scaling2} doesn't hold, so that $\Pi^{AA}=  \mathcal{C}_1$, $\Pi^{AX}= \epsilon \mathcal{C}_2$, $\Pi^{XX}_{SM}= \epsilon^2 \mathcal{C}_3$ for generically $\sim \mathcal{O}(1)$  differences between the various $\mathcal{C}_i$, then the cancellation of the leading terms in Eq.~\ref{eq:cancel} does not occur, and the generic scaling $\Gamma\sim T$ does hold.  
Thus one expects the cosmological production rate of a kinetically-mixed dark photon in the early universe to be parametrically different from that of a $B-L$ gauge boson, which has a distinct coupling structure from the photon.   Another interesting case is a dark Higgs boson $S$ \cite{Krnjaic:2017tio}, which can have unsuppressed thermal production in two ways. The interaction Lagrangian coupling $S$ to the SM Higgs does directly give $S$ unique couplings to the SM plasma through its interactions with the Higgs.  Once the Higgs boson leaves the plasma shortly after electroweak symmetry breaking, the dark Higgs inherits all of its couplings to species remaining in the SM plasma from mixing with $H$; however,  as the Higgs boson itself is gone,  the SM production of $S$ can still be unsuppressed.

Our focus in this paper will be on the case where the leading thermal scattering rates between HS and SM particles are unsuppressed in the $m/T\to 0$ limit.  This is partly for computational tractability, as it makes kinetic theory calculations a reliable guide to the temperature-dependence of the theory, and partly because these scenarios allow us to reveal some novel cosmological behavior.  In these models, the production of dark states from the SM is dominated by
 $2\to 2$ scattering, e.g.~$f g\to X f$, where, for example, $X$ may be a $B-L$ dark vector boson or a Higgs-mixed dark scalar.   Importantly, this particle $X$ is not the dark matter, and is typically unstable on cosmological time-scales.  These $2\to 2$
processes have scattering rates that generically scale as $\Gamma\sim n\langle
\sigma v \rangle \propto T$, and dominate the interactions between sectors when $T\gtrsim m_{X}$ \cite{Evans:2017kti}. The collision term $C_E$ describing the energy transferred between the two sectors through these scattering processes is given by the
thermal average of the scattering amplitude weighted by the energy
carried by the dark particle.  If SM particles 1 and 2 scatter to SM particle 3 and a dark particle, labeled 4, the 
collision term can be expressed as
\barray
\nonumber
\label{eq:coll}
C_E &=& \int d\Pi_i ( 2\pi) ^ 4\delta ^ 4 (\sum p_i) E_4
|\mathcal{M}(12\to 34)|^2\\
  &&\phantom{move me over}\times \left(f_1 f_2 (1\pm f_3)(1\pm f_4) - f_3 f_4 (1\pm f_1)(1\pm f_2) \right)
  \label{eq:CollisionTerm}\\
  \nonumber
  &\equiv& C_E^f - C_E^b .
\earray
Here in the last line we have  introduced separate notation for the collision term governing forward scattering, $C_E^f$, which deposits energy into the dark sector, and the  backward scattering term $C_E^b$, which transfers energy from the hidden sector back to the SM.
In evaluating these collision terms, we will use classical (Maxwell-Boltzmann) statistics
for simplicity.  As these collision terms are important at energies where $T\gg m_i$ for all particles involved, a priori the full dependence on quantum statistics should be retained.   Fortunately, dark mediator production from the SM thermal bath is typically dominated by semi-fermionic processes such as $f g\to X f$, for which empirically we find  that  Maxwell-Boltzmann statistics provide a reasonable approximation to the full result, accurate to within a factor of $\lesssim 2$ (see also \cite{Adshead:2016xxj, Evans:2017kti, Adshead:2019uwj}).   

\begin{figure}
\begin{center}
\includegraphics[width=0.8\textwidth]{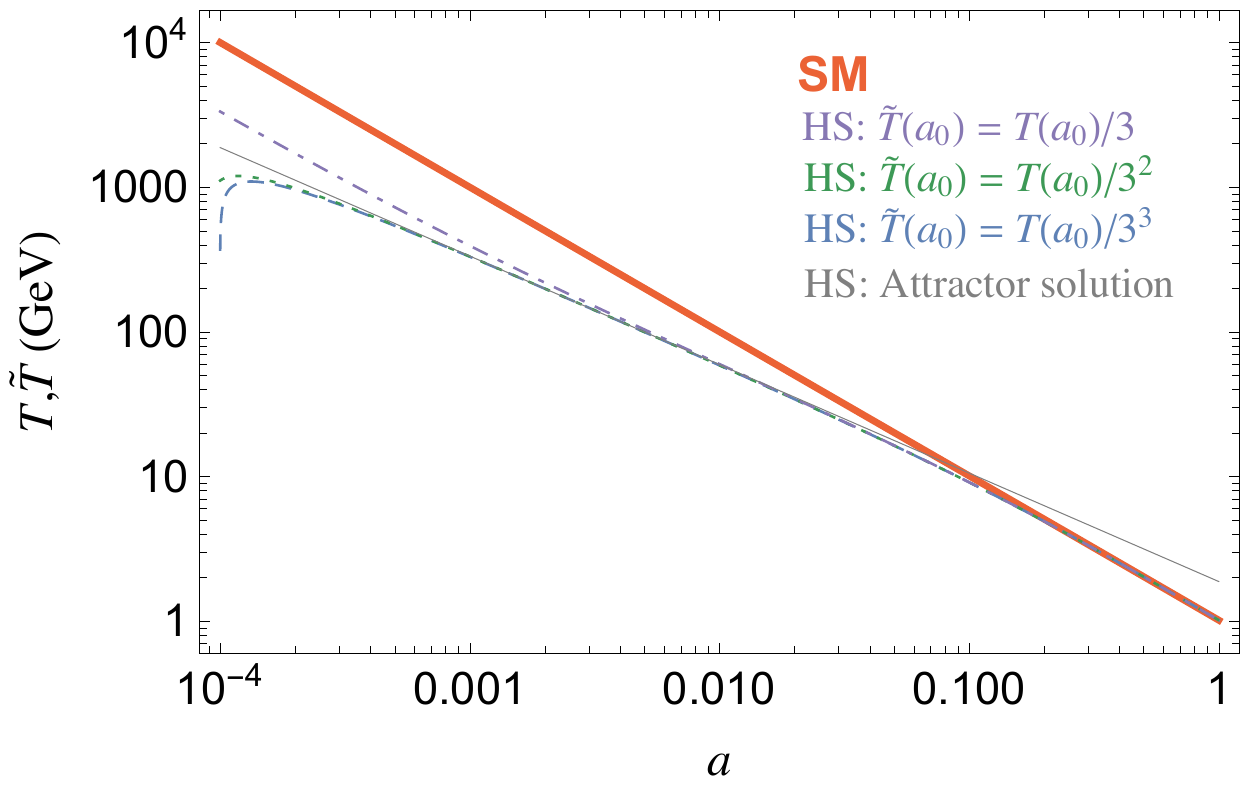}
\caption{Evolution of  SM temperature (red) and  hidden sector
  temperature 
  (purple, green, blue) as a function of scale factor $a$
  when the two sectors are linked by the renormalizable $2\rightarrow 2$
  interaction of Eqs.~\ref{eq:cef} - \ref{eq:ceb}.  Different HS temperature solutions follow from
  different initial conditions at
  $a_0$. The grey line shows the attractor `leak-in'
  solution, $ \tilde T\propto a^{-3/4}$, of Eq.~\ref{eq:THSofTSM}. 
  Solutions with initial conditions below the leak-in attractor rapidly
  converge to it, while the solution that starts at a higher temperature 
  than the leak-in solution  redshifts until it matches onto the attractor.
  The two sectors equilibrate near $a = 0.1$.}
\label{fig:toymodel1}
\end{center}
\end{figure}

To gain some quantitative intuition for the behavior of a leak-in sector, 
consider a toy model where  the leading process transferring energy 
between the SM and HS radiation baths is described by a constant matrix element
$\mathcal{M}=\epsilon$, neglecting all particle masses.  In this case, the
collision term describing forward energy transfer 
takes the simple form
\beq
\label{eq:cef}
C_E^f(T) = \frac{\epsilon^2}{64 \pi^5} T^5,
\eeq
while the backward energy transfer from the reverse process is well-approximated as 
\beq
\label{eq:ceb}
C_E^b(T,\tilde T) =  \frac{\epsilon^2}{64 \pi^5} T^2 \tilde T^3.
\eeq
The resulting temperature evolution for both sectors is shown in
Fig.~\ref{fig:toymodel1}, in the approximation that the SM dominates
the Hubble expansion.  Before the two sectors equilibrate, the hidden
sector temperature exhibits a characteristic `leak-in' phase,
which realizes a quasi-static equilibrium  between the energy injection from the SM and
the energy dilution from the expansion of the universe. Hidden sectors that
have a small initial reheat temperature rapidly rise up to reach the
leak-in solution, as seen in the 
green and blue 
curves in Fig.~\ref{fig:toymodel1}.  Meanwhile, if the hidden sector has a
reheat temperature higher than the temperature of the leak-in phase, as for the 
purple 
curve in Fig.~\ref{fig:toymodel1}, it redshifts like a standard
adiabatic radiation bath ($\tilde T\propto 1/a$) until its temperature
reaches the leak-in solution, at which point the energy injection from
the SM is no longer negligible.  The leak-in phase is thus an attractor
solution, and in particular, at any given value of the SM temperature,
the hidden sector temperature during leak-in is completely determined by the resulting 
energy transfer rate.  This cosmology is thus IR-dominated, i.e.,
once the leak-in phase is attained, there is no remaining dependence
on the initial conditions in the hidden sector.  This ensures that the
properties of DM freezing out during the leak-in phase are independent
of $T_{RH}$.

\subsection{Essential properties of the leak-in phase}
\label{sec:analyticLI}

We can obtain several useful properties of the leak-in phase by analytically solving 
Eqs.~\ref{eq:b1}--\ref{eq:b3} in the regime where $\tilde T \ll T$, and therefore 
\begin{itemize}
\item the energy of the universe is dominated by the SM radiation bath, $H\propto
T^2/M_P$;
\item we can neglect the backward energy transfer rate into the SM; and
\item the SM entropy is approximately conserved,
$T\propto 1/a$.
\end{itemize}
With these assumptions, the evolution of the hidden sector energy
density is given simply by
\beq
\label{eq:simpleboltz}
\dot{\tilde\rho}+4H \tilde\rho =C_E^f(T).
\eeq
Let us now take $C^f_E = c_E T^5$, where $c_E$ is a numerical constant, as dimensional analysis requires when all masses are negligible.  This will be a good approximation to the collision term in realistic models away from mass thresholds. In the toy model of Eqs.~\ref{eq:cef} and~\ref{eq:ceb},  $c_E = \epsilon^2/(64\pi^5)$.
With $\tilde \rho = (\pi^2/30) \tilde g_{*}(\tilde T) \tilde T^4$, Eq.~\ref{eq:simpleboltz} can be easily solved to obtain 
\begin{equation}
\label{eq:THSofTSM} 
\tilde T^4 =   \sqrt{\frac {45}{ 4\pi^3 g_{*}}} \,\frac{30}{\pi^2\,  \tilde g_{*}}   \,              c_E M_{Pl}  T^3.
\end{equation} 
This expression for the HS temperature lets us observe two important things.
Firstly, $\tilde T\propto T^{3/4}\propto a^{-3/4}$---the HS radiation bath redshifts as if
it were matter, and in particular dilutes less slowly than an adiabatic radiation bath.
Secondly, the HS temperature is completely determined by the SM temperature and the strength 
of the leak-in interaction $c_E$, so that it scales with the small portal coupling as $\tilde T \propto c_E^{1/4} \propto \epsilon
^{1/2}$.  

It is worth noting that the scaling $\tilde T \propto \epsilon^{1/2}$ requires only that $C_E(T,\tilde T) \approx C^f_E(T)$, i.e., it does not
depend on the specific functional dependence on the SM temperature $T$, but is a direct consequence of taking
the hidden sector cold compared to the SM. When $\tilde T \ll T$, the Boltzmann equation 
describing the hidden sector evolution, Eq.~\ref{eq:b2}, can be written as
\beq
\frac{d\tilde T}{da} = \left( \frac{C_E^f(T)}{H(T) \frac{2 \pi^2}{15}\,\tilde g_*\,\tilde T^4 }-1\right) \frac{\tilde T}{a},
\eeq
where both $C_E^f$ and $H$ are functions of the SM temperature only in this limit.
But then, as $C_E^f \propto \epsilon^2$, it is clear that all $\epsilon$ dependence can be scaled 
out by sending $\tilde T \to \tilde T/\sqrt{\epsilon}$.

\subsection{Leak-in, freezeout}
\label{sec:fo}

We would now like to understand what happens to DM that freezes out of
a hidden sector radiation bath during a leak-in phase.  As the leak-in phase is an attractor solution, freezeout during leak-in is a generic possibility, and does not require any fine-tuning of mass scales.
We will begin with some analytic estimates to establish the main features of dark matter freezeout from a leak-in phase --- or, for short, ``leak-in dark matter'' (LIDM) 
and highlight how it differs from a thermal relic in an adiabatic hidden sector.

As a warm-up, we begin with a reminder of DM freezeout in a decoupled, but adiabatic, hidden
sector \cite{Feng:2008mu}, i.e., hidden sector freezeout where $\tilde T=\xi T$ for a constant $\xi $.
The sudden
freezeout approximation, $n(\tilde x_f)\langle\sigma
v\rangle = H(x_f)$, lets us estimate the DM relic abundance as a
function of its annihilation cross-section.  Here we have defined $\tilde x = m/\tilde
T$ and $x = m/T$, with $m$ the DM mass.  Assuming that the SM energy dominates Hubble, the sudden freezeout approximation implies
\begin{equation} 
e^{\tilde x_f} =\left(\frac{d_\chi m^3\langle \sigma v\rangle}{ (2\pi) ^
    {3/2}H(m)}\xi^2\right)\tilde x_f^ {1/2}\equiv A \tilde x_f^ {1/2},
\label{eq:smfo}
\end{equation} 
where $H(m) $ is evaluated at $x = 1$ and $d_\chi$ is the number of degrees of freedom for the dark matter.   Iteratively solving this equation for $\tilde x_f$
yields the approximate solution $\tilde x_f = \ln A
+\frac{1}{2}\ln\ln A$.  To facilitate comparison with the
canonical WIMP it is convenient to consider the yield $Y_\infty\approx 
n(\tilde x_f)/s(x_{f})$,
where $x_f\equiv \xi \tilde x_f$ is the value of the SM temperature at DM freezeout,
\begin{equation} 
Y_\infty =   \sqrt\frac{45}{\pi} \frac{g_*^{1/2}}{g_{*S}}  \frac{\xi \tilde x_f}{m M_{Pl}\langle \sigma v\rangle}.
\end{equation} 
Necessarily this reduces
to the standard result when $\xi\to 1$. To
leading order, obtaining the correct relic abundance for DM freezing out in a cold adiabatic HS requires the annihilation cross-section to be rescaled as $\sv \to \xi \sv$, as the
dependence of $\tilde x_f$ on $\xi$ is only logarithmic.

Now let us repeat this exercise for DM freezing out of a leak-in radiation bath.
In this case, we can read off from Eq.~\ref{eq:THSofTSM} that $\tilde x$ is related to $x$ through
\begin{equation}
x =\left[c_E\times \frac{\mpl}{m}  \times  \frac{15\sqrt{45}}{ \pi^{7/2}\sqrt{g_*}\,  \tilde g_{*}}\right]^{1/3} \tilde x^{4/3} \equiv b \tilde x^{4/3},
    \label{eq:xasxt}
\end{equation} 
which lets us express the (SM-dominated) Hubble rate in terms of $\tilde x$,
\begin{equation} 
H(\tilde x) = \frac {H(m)} {b^ 2\tilde x^ {8/3}}.
\end{equation} 
The sudden freezeout condition for $\tilde x_f$ is then 
\begin{equation} 
e^{\tilde x_f} =\left(\frac{d_\chi m^3\langle \sigma v\rangle}{ (2\pi) ^
    {3/2}H(m)}b^2\right)\tilde x_f^ {7/6}\equiv  A \tilde x_f^ {7/6},
    \label{eq:transcendental}
\end{equation} 
which has the approximate solution $\tilde x_f = \ln \tilde  A
+\frac{7}{6}\ln\ln \tilde A$.  Comparing Eqs.~\ref{eq:smfo} and~\ref{eq:transcendental} we can recognize that $b$ is serving as a ``coldness''
parameter analogously to a fixed constant $\xi$, while the different fractional power of $\tilde x_f$
reflects the different temperature evolution with redshift.  However, $b$ is {\em not} given by the temperature ratio between the two sectors at freezeout, which is rather $\xi(\tilde x_f) = b \tilde x_f^{1/3}$. 

Using Eq.~\ref{eq:transcendental}, we can derive the yield
\begin{equation} 
Y_\infty = Y(x_f,\tilde x_f) =  \sqrt\frac{45}{\pi} \frac{g_*^{1/2}}{g_{*S}} \frac{b \tilde x_f^{4/3}}{m M_{Pl}\langle \sigma v\rangle}
\end{equation} 
(recall that the SM entropy is approximately conserved during the
leak-in phase).  
Thus we can again parametrically expect 
\begin{equation} 
\label{eq:scaling} 
\sv = b \tilde x_f^{1/3} \sv_{W} = \xi(\tilde x_f)  \sv_{W} 
\end{equation} 
where $\sv_W$ denotes the annihilation cross-section for a standard thermal WIMP. 
In particular note the  annihilation
cross-section necessary to obtain the desired relic abundance scales with $\epsilon$ as $b\propto \epsilon^{2/3}$.  
Since here DM freezeout depends on the hidden sector temperature, we typically expect $\tilde x_f \sim 15$, while $x_f \ll 1$ is possible.

\subsection{Region of interest for the portal coupling}
\label{sec:absfloor}

The LIDM mechanism can account for the observed DM
abundance for a bounded range of portal couplings $\epsilon$.  At
sufficiently large values of the dimensionless coupling $\epsilon$, the dark radiation bath will thermalize
with the SM, yielding a WIMP next door \cite{Evans:2017kti}. 
We refer to this transition as the ``equilibration floor.'' For
small enough $\epsilon$, however, the dark sector never reaches a high enough
co-moving dark matter number density 
 to account for the observed DM relic abundance.
This ``absolute coupling floor'' for leak-in dark matter can be
straightforwardly estimated by requiring that the maximum value
attained by the equilibrium leak-in DM yield
\beq
Y_{eq} = \frac { d_\chi } {( 2\pi) ^ {3/ 2} } (m  \tilde T )^{3/2} e^{ 
  - m / \tilde T } \frac { 45} { 2\pi ^ { 2} } \frac { 1} { g _ { *,S } T ^ { 3} }
\eeq
should equal the observed relic abundance, 
\beq
\label{eq:yinf}
Y_\infty =
\frac{\Omega_{DM}\rho_{c,0}}{m s_0} ,
\eeq
where $\Omega_{DM} \rho_{c,0}$ is the present-day energy density of DM
and $s_0$ is the present-day entropy of the CMB.  Using $\tilde
T\propto a^{-3/4}$ and $T\propto a^{-1}$, maximizing $Y_{eq} (a)$ with
respect to $a$ tells us that the maximum yield is obtained at
\beq
\tilde T (a_{max}) =\frac{2 m}{5}.
\eeq
Using Eq.~\ref{eq:xasxt}, the maximum equilibrium yield obtained is then
\beq
Y_{eq}(a_{max}) = 0.21 \times \frac{d_\chi }{\tilde g_{*}}\frac{c_E}{g_{*,S}\sqrt{g_*}}\frac{\mpl}{m}.
\eeq
 Requiring that this maximum yield is greater or equal to the observed relic abundance, Eq.~\ref{eq:yinf},
  places a condition on the strength of the interaction with the SM, 
\beq
c_E \gtrsim 2 \times 10^{-25} \lp \frac{g_{*,S}\sqrt{g_{*}}}{(106.75)^{3/2}} \frac{\tilde g_*}{d_\chi} \rp.
\label{eq:cE}
\eeq
If $c_E$ is below this critical value, then even if the 
sector were to internally thermalize, there would never 
have been a large enough dark matter number density 
to correspond to the observed relic abundance today.  
Recalling $c_E \propto \epsilon^2$, we immediately observe that
the absolute minimum value of $\epsilon$ consistent with the leak-in
scenario is independent of the DM mass (although logarithmic 
dependence on the mass may enter through the collision term).  
This requirement defines an absolute coupling floor, 
below which leak-in cannot produce enough dark matter 
to reproduce the observed relic abundance. 

Of course, within any given model,  the portal coupling $\epsilon$ will be subject to many terrestrial, astrophysical, and cosmological constraints
that depend on the specific properties of the mediator $X$.  Cosmological constraints on the mass and lifetime of $X$ arise due to the relic hidden sector radiation bath in the early universe, which can lead to constraints  through either  its gravitational imprint on the early universe or the decays of $X$ into the SM.  

It is worth noting that these cosmological constraints have some model-dependence, even under the assumption that $X$ is the lightest species in the dark sector.  
In the absence of other dark states that $X$ can interact with, cosmological constraints on $\epsilon$ are  dominated
by ``freezein'' constraints on $X$,
i.e., constraints on the out-of-equilibrium population of mediators produced in the early
universe thanks to their couplings to the SM.  This population is
dominated by the production of $X$s at SM temperatures $T\sim m_X$.  
However, when $X$ is part of a larger dark sector that was once in internal thermal equilibrium, there is a  separate population of $X$s resulting from the relic radiation bath. After the HS bath leaves equilibrium, the freeze-in population will not be able to equilibrate with the relic bath population.  The hidden radiation bath may leave equilibrium long before the late-time injection of freezein $X$, or---depending on the dark particle content---possibly not until freezein has effectively stopped.  These two different scenarios lead to two very different phase space distributions for the final $X$ population, and thus to different potential signatures. In minimal models, such as the one we will discuss below, the number density  of $X$ in a relic bath population can easily exceed the number density in the freeze-in population, and therefore may dominate any constraints arising from the decays of $X$.

This is one example of a general theme: once we depart from thermal equilibrium, the details of which processes go out of equilibrium first can lead to rich behavior even within a simple model, e.g.~\cite{Cheung:2010gj,Chu:2011be,Chu:2013jja,Bernal:2015ova,Heikinheimo:2017ofk,Duch:2017khv,Krnjaic:2017tio,Heikinheimo:2018duk,Heeba:2018wtf,Berger:2018xyd,Forestell:2018dnu}. 
To go further and work out the observational consequences for leak-in DM, we will need to be more concrete and specify a model.
In the next section, we build on this discussion, 
extending this toy model of leak-in DM to a more complete picture of dark matter production in a specific out-of-equilibrium hidden sector.

\section{Dark matter relic abundance in an out-of-equilibrium hidden sector}
\label{sec:addingfi}

\begin{figure}
\begin{center}
\includegraphics[width=0.8\textwidth]{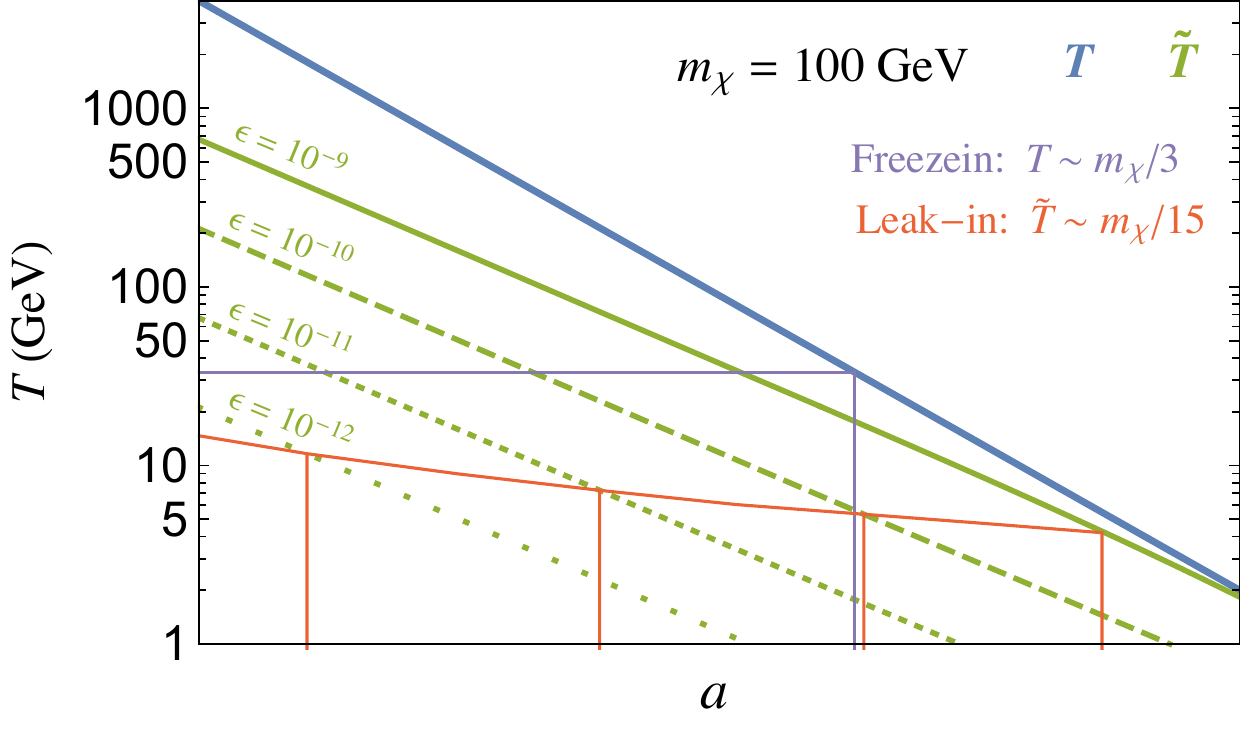}
\caption{Evolution of the SM temperature $T$ (blue) and the HS temperature $\tilde T$ (green) for four choices of the portal coupling $\epsilon$.  Freezein from the SM is dominated by SM temperatures $T\sim m_\chi/3$ (shown in purple), while leak-in DM is dominated by HS temperatures $\tilde T\sim m_\chi/15$ (as indicated by the red curve).  When freezein occurs prior to when LIDM would freezeout (as happens with the solid green line, $\epsilon=10^{-9}$), the injected DM particles can thermalize with the HS plasma, leaving little net contribution to the DM relic abundance.  When freezein occurs after LIDM freezeout, it may have a significant impact on the final relic abundance (as happens with the short dashed line, $\epsilon=10^{-11}$).}
\label{fig:tva}
\end{center}
\end{figure}

In the previous section we developed a general analytic guide to the properties of DM that freezes out during the ``leak-in'' evolution of a hidden radiation bath, which we refer to as leak-in DM.   The same interactions that separately govern leak-in and  freezeout will typically also yield out-of-equilibrium production of DM  directly from the SM, i.e., freezein \cite{McDonald:2001vt, Hall:2009bx}.  Although direct production of DM from the SM will generally give a sub-leading contribution to the total \emph{energy} density of the HS radiation bath, it has the potential to substantially affect the final DM \emph{number} density, and thus the relic abundance. The leak-in mechanism is controlled by the HS temperature, and is governed by the properties of the HS radiation bath at $\tilde T \sim m/15$.  Freezein production, on the other hand, is dominated by SM temperatures near the dark matter mass, $x_{fi}\sim 3-5$.  Depending on the  coldness of the HS relative to the SM, direct production of DM from the SM may thus dominantly occur either prior to HS freezeout, i.e., $x_{fi} < x_f$, in which case its ultimate impact is negligible, or post-HS freezeout, $x_{fi}>x_f$, in which case it can sometimes, but not always, dominate the final DM abundance (see Fig.~\ref{fig:tva}).

In this section, we introduce a specific model of a minimal hidden sector for concreteness, consisting of Dirac fermion dark matter $\chi$ coupled to a (massive) dark vector $Z_D$ that couples to the SM through $B-L$ charges.  We detail the resulting interplay of leak-in, freezein, and ``reannihilation'' \cite{Chu:2011be} in determining the DM relic abundance when the coupling between $Z_D$ and the SM fields via the $B-L$ interaction is too small to allow the dark sector to achieve equilibrium with the SM.  We assume throughout that the dark  sector is internally thermalized; criteria for attaining internal thermalization are discussed in Appendix~\ref{sec:inttherm}. 

\subsection{A minimal $B-L$ vector portal leak-in hidden sector}
\label{sec:model}

We consider a minimal hidden sector consisting of a Dirac fermion
DM candidate, $\chi$, and a massive dark vector, $Z_D$.  
This dark vector is the gauge boson for a 
$U(1)$ symmetry, and interacts with the SM by coupling to the $B-L$ current \cite{Marshak:1979fm,Heeck:2014zfa,Bauer:2018onh} 
\beq
\mathcal L =  g_D  \bar \chi \gamma_\mu \chi Z_D^\mu +  \epsilon \sum_f Q_f  \bar f \gamma_\mu f  Z_D^\mu,
\label{eq:zcouplfinal}
\eeq
where $Q_f$ is $\pm 1$ for leptons and $\pm 1/3$ for quarks. 
 For simplicity and minimality, we consider a St\"uckelberg origin for the dark vector mass \cite{Stueckelberg:1938zz,Feldman:2007wj}.  
Since we are interested in dark sectors that never attain thermal equilibrium with the SM, the $B-L$ portal coupling $\epsilon$ is assumed to be very small.   
This model thus assumes a large hierarchy between the couplings of $Z_D$ to DM and to the SM, $g_D\gg \epsilon$, which, while technically natural, does invite model-building questions.   This hierarchy of couplings could originate from (e.g.) dark matter with a very large $B-L$ charge, or from a $U(1)_{B-L}\times U(1)_D$ symmetry broken at a higher scale.  In principle, UV model-building can introduce some model-dependence through the introduction of new particles in the UV. To insulate the discussion from this UV sensitivity, we will simply take $g_D\gg \epsilon$ throughout the discussion of the next two sections, but in Appendix~\ref{sec:model}, we will provide some simple UV completions to this hierarchical $B-L$ model and discuss their consequences.\footnote{Consistently gauging the SM $B-L$ symmetry does require the introduction of three right-handed neutrinos $N$.  In the model used here, these RH neutrinos never equilibrate with either the SM or the hidden sector plasma.  Accordingly, we neglect RH neutrinos for the purposes of estimating the rates and constraints relevant for LIDM.}

This minimal hidden sector can be described by four independent free parameters,
which we will take to be $\alpha_D,\epsilon,m_\chi$ and $m_{Z_D}$.  So long as $m_\chi \gtrsim 10\, m_{Z_D}$, such that $Z_D$ is relativistic at the time of DM freezeout, 
the DM relic abundance will be largely insensitive to the dark vector mass: both the DM annihilation cross-section (discussed below) and the temperature evolution of the radiation bath prior to and during freezeout are largely independent of the dark vector mass when the dark vector is relativistic.  Throughout this paper, we will thus consider $m_{Z_D} \leq m_\chi/10$ in order for this specific minimal hidden sector to serve as a useful illustration of the dynamics of a general dark sector in a leak-in phase. 

We compute the energy transfer collision term, $C_E$, by considering processes that produce dark vectors from interactions with the SM plasma.  In particular,  we sum up the contributions from $g f \to Z_D f$, $f \bar f \to Z_D g$, $\gamma f \to Z_D f$, $f\bar f \to Z_D \gamma$, over all fermions 
\cite{Evans:2017kti}.
With the collision term in hand, the hidden sector temperature can be  determined numerically.  A particularly useful function is the ratio of hidden sector and SM temperatures, $\xi = \tilde T/T$. If the transfer of energy out of the hidden sector is negligible, $C_E(T,\tilde T)\approx C_E^f(T)$, we have 
\beq
\xi(T) = \lp \int_{T_i}^{T} d\bar T \frac{30 C_E^f(\bar T)}{\pi^2 \tilde g_{*} H(\bar T) \bar T^5} \rp^{\frac 14} \propto \epsilon^{\frac 12}T^{-\frac 14},
\eeq
which exhibits the scaling derived in Sec.~\ref{sec:analyticLI}.
For derivations and further details, including incorporation of the backwards collision term near the equilibration floor, see Appendix \ref{app:CE}.

\subsection{Interplay of leak-in dark matter and freezein processes}
\label{sec:others}

\begin{figure}
\begin{center}
\includegraphics[width=0.8\textwidth]{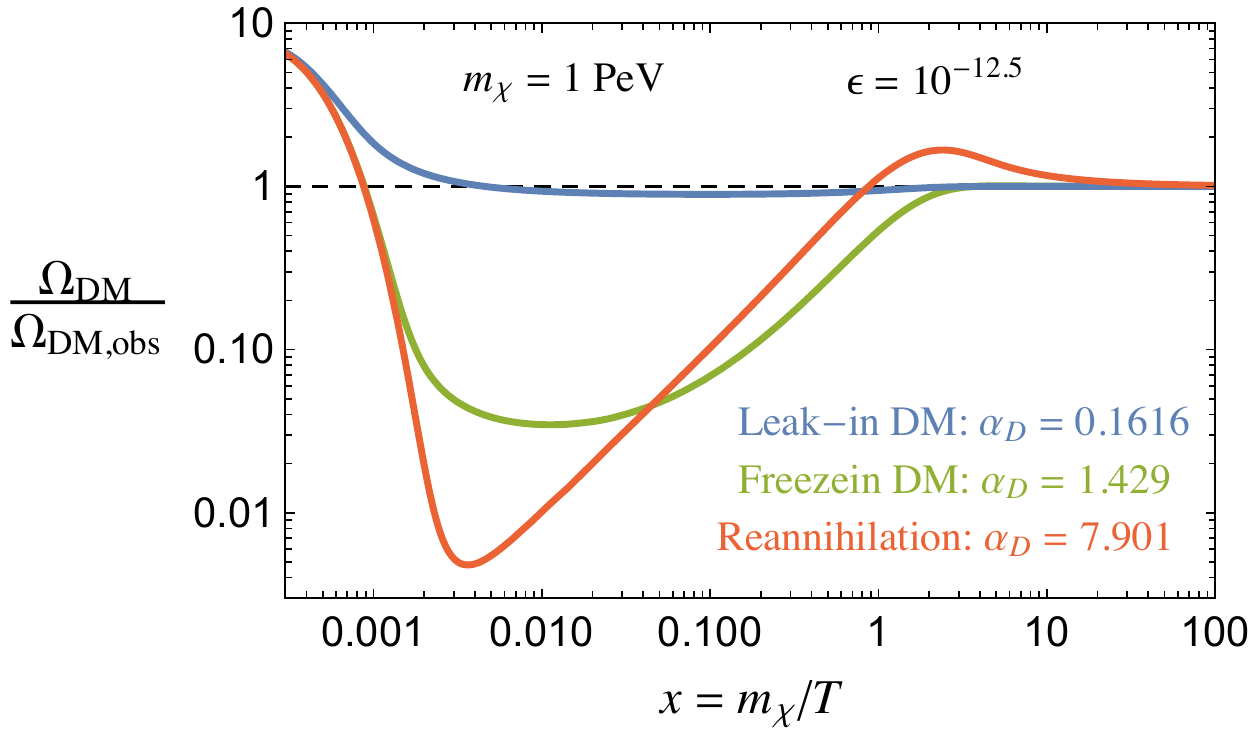}
\caption{Evolution of the DM number density as a function of $m_\chi/T$ for a point with three possible solutions for $\alpha_D$.  The smallest value of $\alpha_D$ (blue) corresponds to a dominantly leak-in solution (with a small contribution from freezein), the intermediate value of $\alpha_D$ (orange) gives a dominantly freezein solution (with a small contribution from leak-in), and the largest $\alpha_D$ (green) yields reannihilation.  The parameters used for the figure were selected to emphasize the difference between the three cases.}
\label{fig:3sol}
\end{center}
\end{figure}

In addition to leak-in,  below the equilibration floor there are two related processes that can govern the relic abundance, freezein \cite{McDonald:2001vt, Hall:2009bx} and reannihilation \cite{Chu:2011be}.  
``Freezein'' refers to an out-of-equilibrium dark matter population injected predominantly near $T\sim m_\chi/(2-5)$ with little subsequent evolution, while ``reannihilation'' occurs when the freezein mechanism injects much more dark matter than is needed, but a large coupling between the DM and a dark mediator allows for the excess to annihilate down to the correct relic abundance, with this depletion typically completing near $T\sim m_\chi/10$.  Example evolution of the DM number abundance with temperature is shown for all three processes in Fig.~\ref{fig:3sol}.   

In this minimal $B-L$ model, dark matter freezeout is governed by the $\chi\bar\chi \to Z_D Z_D$ annihilation process with  cross-section
\beq
\vev{\sigma_{\chi\bar\chi\to Z_DZ_D} v} = \frac{4\pi \alpha_D^2}{m_\chi^2} \frac{(1 - r^2)^{\frac 32}}{(2-r^2)^2} + \mathcal{O}(v^2),
\label{eq:DMann}
\eeq
where $r \equiv m_{Z_D}/m_\chi$ and  $\alpha_D\equiv g_D^2/4\pi$.  Freezein, however, is dominated by the direct production of DM from the SM through $s$-channel $Z_D$ exchange, $ f\bar f\to \chi\bar\chi$:
\beq
\sigma_{ f\bar f\to \chi\bar\chi} = \frac{\eta_f \alpha_D \epsilon^2}{3s^3}  \sqrt{\frac{s-4m^2_{\chi}}{s-4m^2_f}} \lp s+ 2m_f^2\rp\lp s+ 2m_\chi^2\rp
\eeq
where $\eta_f= 1/3$ $(1)$ for quarks (leptons). 
A DM particle produced via freezein will, in the presence of a dark radiation bath at $\tilde T$, rapidly attain kinetic equilibrium in the parameter space of interest, though not necessarily chemical equilibrium. Thus, given $\tilde T$ as a function of $T$, $\tilde T(T;\epsilon)$, we can obtain the relic abundance of DM by solving the single Boltzmann equation \cite{Chu:2011be,Krnjaic:2017tio}
\beq
\frac{dY_{\chi}}{dT} =K(T)\left[ \vev{\sigma_{\chi\bar\chi\to Z_DZ_D} v} \lp Y_{\chi}^2 - Y_{\chi,eq}^2 (\tilde T,T) \rp-\sum_f \vev{\sigma_{ f\bar f\to \chi\bar\chi} v}  Y_{\chi,eq}^2  (T)   \right],
\label{eq:Boltz}
\eeq
where 
\beq 
K(T) \equiv M_{Pl} \sqrt{\frac{\pi}{45}}\frac{g_{*S}(T)}{\sqrt{g_*(T)}},
\eeq
$Y_{\chi,eq} (\tilde T,T) \equiv n_{\chi,eq} (\tilde T)/s(T)$ is the equilibrium number density as dictated by the hidden sector temperature, relative to the SM entropy, which we approximate as conserved,\footnote{For cold hidden sectors, $\tilde T \ll T$, this is an excellent approximation.  The approximation further remains reasonable even near the equilibration floor, provided $\tilde g_{*S}(\tilde T)\ll g_{*S}(T)$, which holds in the phenomenologically viable portions of the parameter space.}  and $Y_{\chi,eq} (T)\equiv Y_{\chi,eq} (T,T)$.

\begin{figure}
\begin{center}
\includegraphics[width=0.8\textwidth]{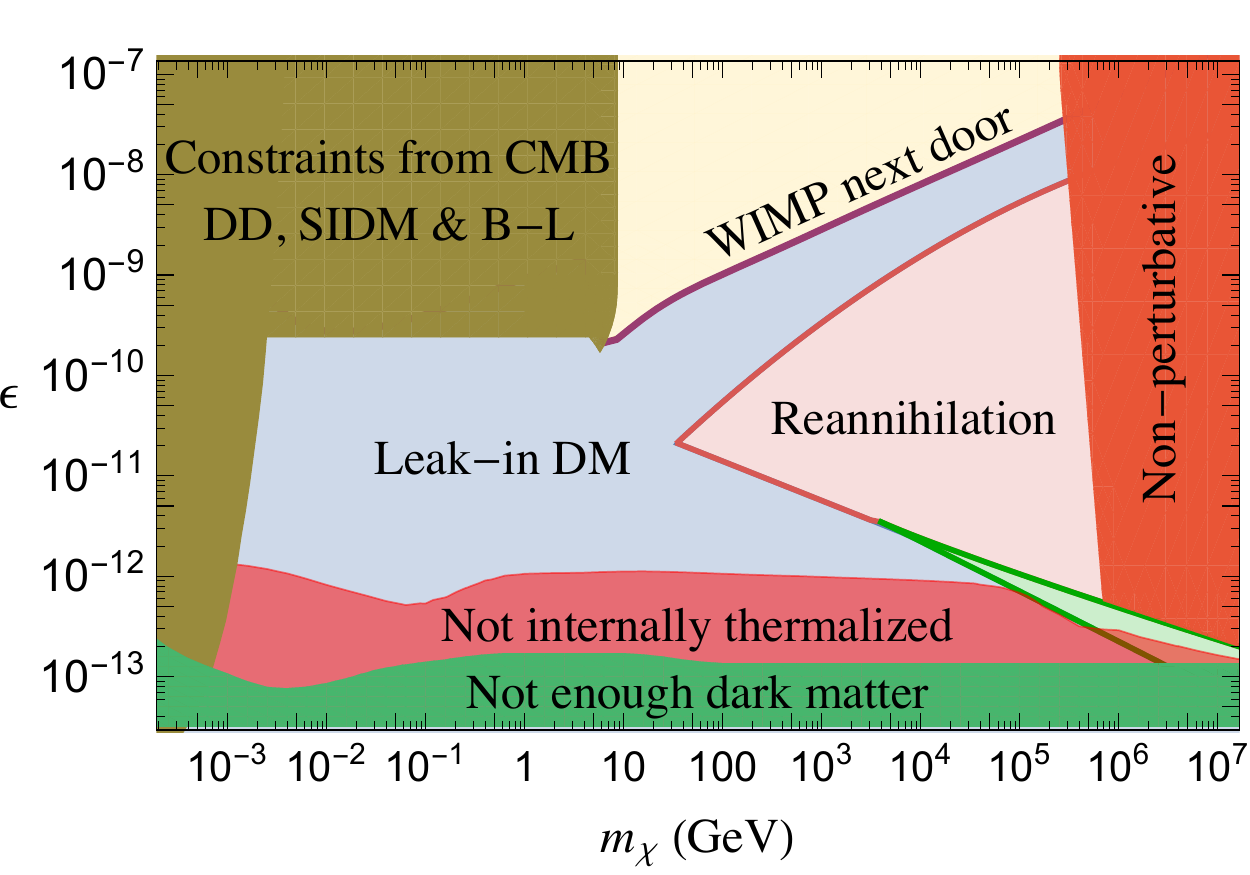}
\caption{The different regions in the parameter space.  Leak-in dark matter is shown in light blue, with the narrow slice near the equilibration floor (purple line) corresponding to late LIDM, and the smaller epsilon region corresponding to early LIDM.  Above the equilibration floor is the WIMP next door \cite{Evans:2017kti}.  Reannihilation \cite{Chu:2011be} is shown with shaded light red.  The narrow slice of parameter space above $m_\chi \sim 3$ TeV in shaded green indicates the three-solution region where three different choices of $\alpha_D$ can produce a mostly LIDM, freezein, or reannihilation solution (see Fig.~\ref{fig:3sol}).   At high masses, the solid red denotes where $\alpha_D >4\pi$ and non-perturbative couplings are clearly required to produce the right relic abundance.  At very small $\epsilon$, the model is below the absolute coupling floor, and not enough dark matter is produced to reach the relic abundance (dark green).  The dark red shaded region above this indicates where the internal thermalization conditions are not satisfied.  At small masses, constraints from the CMB, direct detection, and DM self-interactions together with constraints on the $B-L$ boson forbid any valid DM solutions (shown in olive), as will be discussed in Sec.~\ref{sec:signals}. }
\label{fig:montanaroadmap}
\end{center}
\end{figure}

LIDM is realized when the second  term in Eq.~\ref{eq:Boltz} is unimportant for determining the final relic abundance, i.e., neglecting the effect of that source term will have only small effects on the final dark matter population.  This can happen in two separate regimes.  The first regime occurs when the hidden sector temperature is relatively close to the SM temperature, such that DM produced by freezein can reach thermal equilibrium with the dark radiation bath prior to freezeout (see  Fig.~\ref{fig:tva}): we call this ``late'' LIDM.  The second regime occurs at very small values of $\epsilon$ and $\alpha_D$, where freezeout occurs before freezein stops, but freezein processes are sufficiently feeble to contribute only a tiny fraction to the final DM abundance. 
We call this more weakly coupled regime ``early'' LIDM.   
When the second term in Eq.~\ref{eq:Boltz} is not negligible, we find that generically the DM relic abundance is obtained through reannihilation.  Freezein occurs when the first term is entirely negligible in comparison to the second term, and is realized in a very limited fraction of parameter space.

Fig.~\ref{fig:montanaroadmap} shows a schematic of the viable parameter space and of the mechanisms yielding the correct DM relic abundance in the minimal $B-L$ model.  At large portal couplings above the purple line, the HS and SM sectors are in thermal equilibrium, yielding a WIMP next door scenario  \cite{Evans:2017kti}.    At small portal couplings, the co-moving number density of dark matter is never high enough to produce the correct relic abundance (\ref{eq:cE}).   In practice, the high multiplicity of the SM sector and size of $\alpha_s$ result in $C_E\sim \{\text{few}\}\times \epsilon^2T^5$, placing the absolute coupling floor near $\epsilon\sim 10^{-13}$.  For values of $\epsilon$ slightly above this floor, the hidden sector does not attain internal thermal equilibrium (for any $m_{Z_D}$).  While we will assume internal thermal equilibrium in this subsection, we will establish the validity of this assumption in Appendix~\ref{sec:inttherm}.  At high DM masses, the requisite $\alpha_D$ becomes non-perturbative, while small DM masses are excluded by a combination of constraints on $B-L$ vector bosons, CMB distortions, and DM self-interactions, as detailed below in Sec.~\ref{sec:signals}.

  Late LIDM governs the DM abundance near the equilibration floor, while early LIDM governs the region of parameter space at small $\epsilon$, and the two regimes transition smoothly into each other at smaller values of the DM mass. At larger masses and intermediate values of $\epsilon$, re-annihilation governs the DM abundance.  The boundary between reannihilation and late LIDM is set by requiring that the coupling as determined in a ``leak-in-only'' solution (i.e., $\vev{\sigma_{ f\bar f\to \chi\bar\chi} v}\to0$ in Eq.~\ref{eq:Boltz}) differs from the full solution by less than 10\%.  The boundary between reannihilation and early LIDM occurs in practice when the value of $\alpha_D$, $\epsilon$, and $m_\chi$ cross the point where a ``freezein-only'' solution (i.e., $\vev{\sigma_{\chi\bar\chi\to Z_DZ_D} v}\to0$ in Eq.~\ref{eq:Boltz}) would produce the observed relic abundance.     
There is a narrow slice of parameter space at high DM mass and small $\epsilon$, shown here in green, where a dominantly leak-in, a dominantly freezein, or a reannihilation solution can be achieved for different choices of $\alpha_D$.  One such point is shown in Fig.~\ref{fig:3sol}.  To distinguish between dominantly leak-in and dominantly freezein solutions, we consider the co-moving number density immediately after leak-in freezeout occurs, and ask whether it is greater (leak-in) or less (freezein) than 50\% of the observed value.  In all cases, near transitions both source terms in Eq.~\ref{eq:Boltz} are important for obtaining the final DM abundance.

\begin{figure}
\begin{center}
\includegraphics[width=0.49\textwidth]{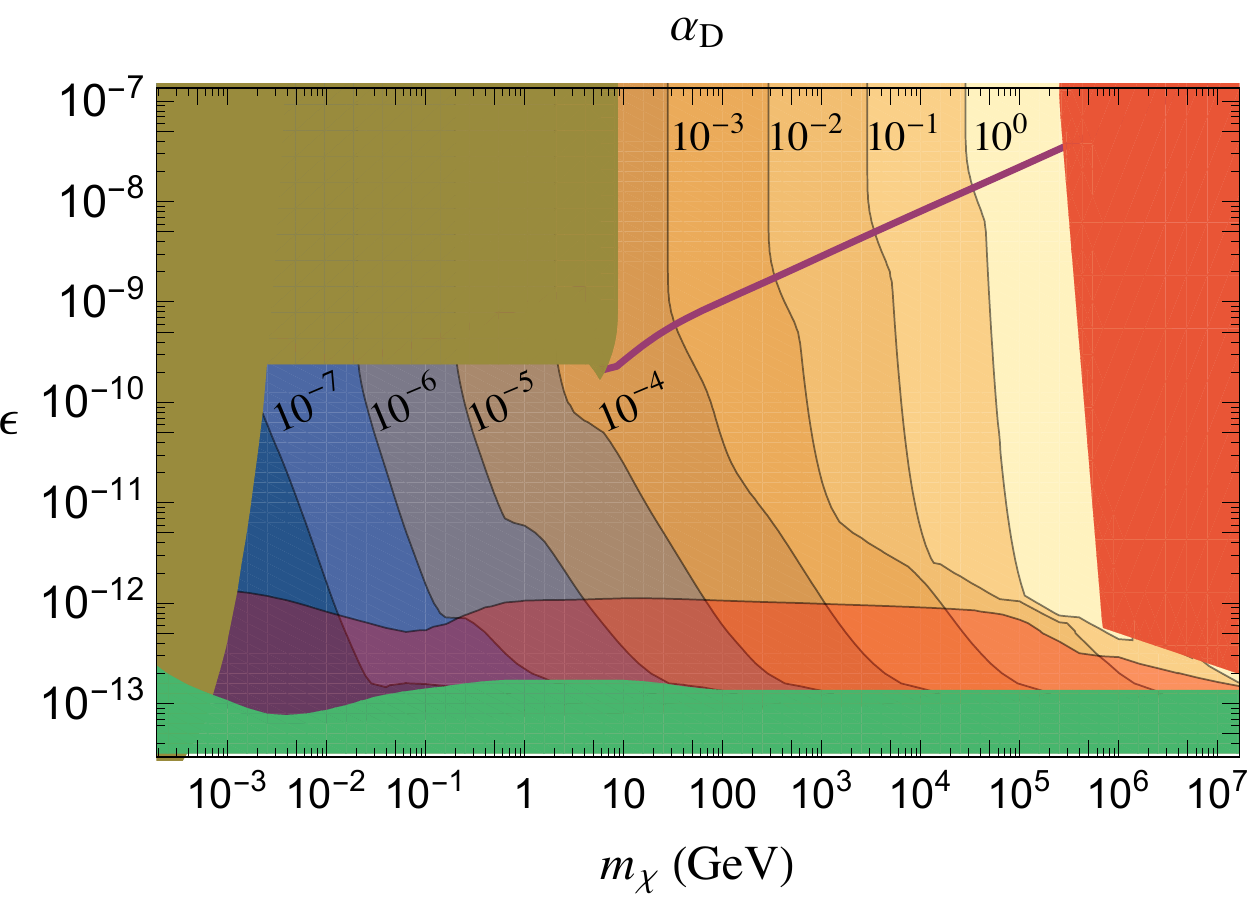}
\includegraphics[width=0.49\textwidth]{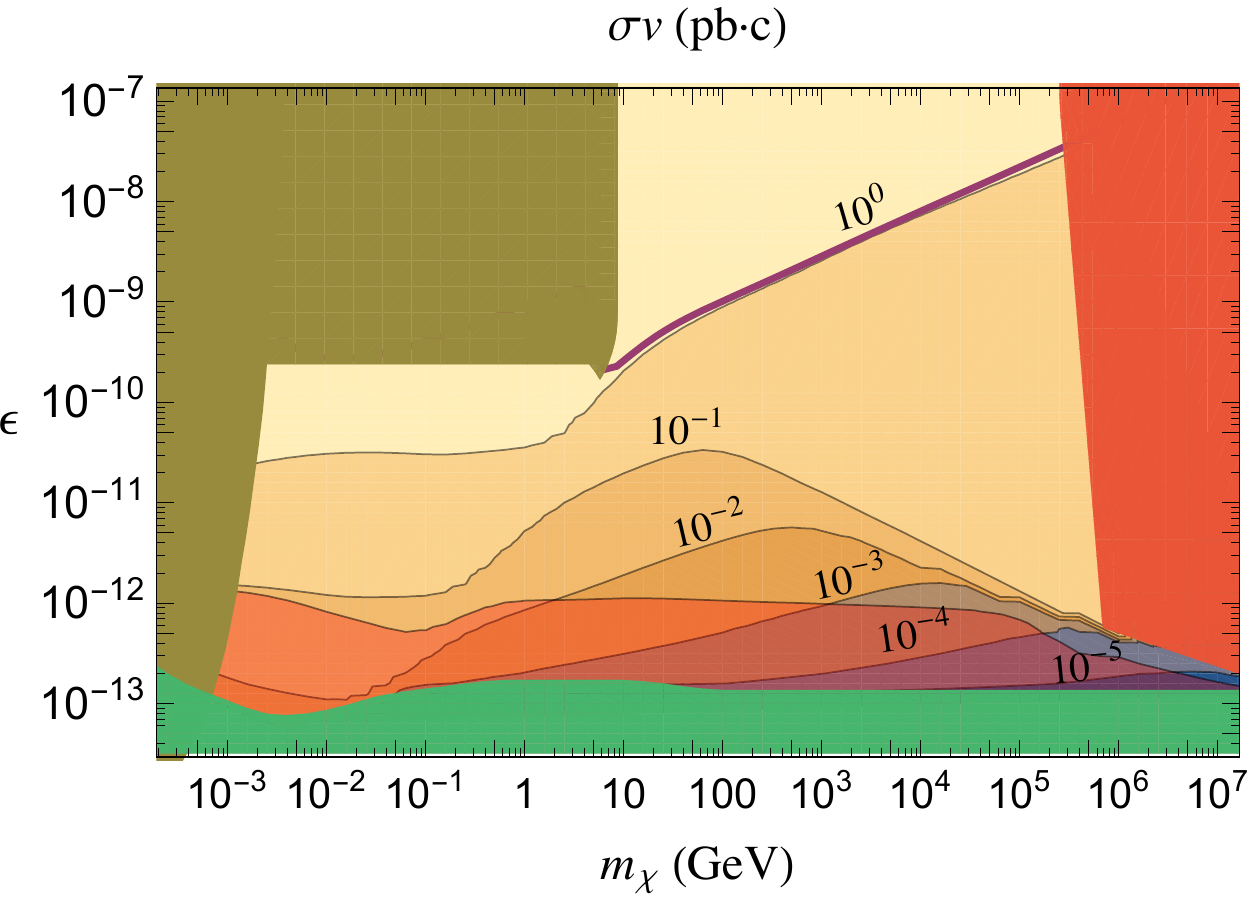}
\caption{{\bf Left:} Values of $\alpha_D$ that produce the correct relic abundance.  {\bf Right:} The corresponding annihilation cross-section (in pb$\cdot$c).  On each plot the equilibration floor is shown by the purple line.  At high masses, the annihilation becomes non-perturbative.  At low masses, various constraints from the CMB, self-interactions, and direct constraints on  $B-L$ vectors forbid any valid dark matter solutions (olive) as will be discussed in Sec.~\ref{sec:signals}.  At small $\epsilon$ the model encounters the absolute coupling floor  (dark green), below which the dark matter abundance is never large enough to produce the observed relic density.  Above this, in shaded red the sector does not satisfy our internal thermalization conditions (Appendix~\ref{sec:inttherm}).}
\label{fig:alphaDsigma}
\end{center}
\end{figure}

For a given $m_\chi$ and $\epsilon$, there is usually a unique value of $\alpha_D$ that realizes the correct DM relic abundance, shown in the left panel of Fig.~\ref{fig:alphaDsigma}.  
Within the three-solution region, we display the $\alpha$ and $\vev{\sigma v}$ values for the mostly leak-in solution.  The corresponding annihilation cross-sections are displayed in the right panel of Fig.~\ref{fig:alphaDsigma}, where the wedge of the reannihilation region is clearly visible at high DM mass.  In the absence of the freeze-in term in Eq.~\ref{eq:Boltz}, the annihilation cross-section would display the simple scaling with $\epsilon$ expected from Eq.~\ref{eq:scaling}.  However, as Fig.~\ref{fig:alphaDsigma} shows, the presence of the freeze-in term instead leads to reannihilation and its larger annihilation cross-sections controlling the phenomenology.  The net annihilation cross-sections are thus only slightly suppressed compared to expectations for a traditional WIMP over much of parameter space, with correspondingly better prospects for detectability; of course, DM in this model can also be much heavier than a traditional WIMP.   At lower DM masses, where the leak-in solution dominates, the numerous mass thresholds of the SM obscure the scaling of Eq.~\ref{eq:scaling}.

\begin{figure}
\begin{center}
\includegraphics[width=0.8\textwidth]{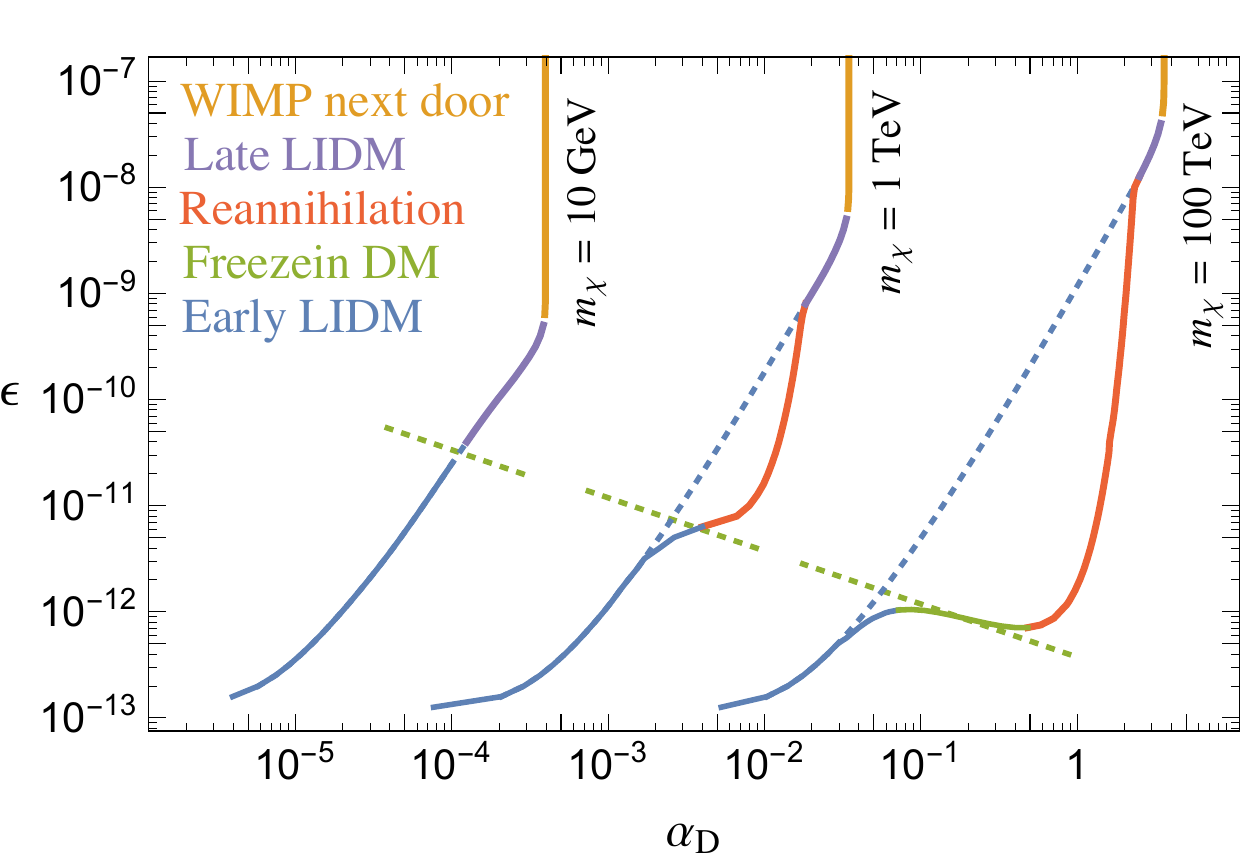}
\caption{The breakdown of correct relic abundance solutions in the $(\alpha_D,\epsilon)$ parameter space for three different dark matter masses (1 GeV, 100 GeV, 10 TeV).  Hidden sector freezeout while in thermal equilibrium with the SM (WIMP next door) is shown in orange.  Late LIDM is shown in purple.  Reannihilation is shown in red.  Freezein is shown in solid green. Early LIDM is shown in solid blue.   The ``LIDM-only'' solution, obtained by neglecting the freezein population from $f\bar f\to\chi\bar\chi$ processes, is shown with the dashed blue line.  The ``freezein-only'' solution obtained by considering only the $f\bar f\to\chi\bar\chi$ and neglecting $\chi\bar\chi\to Z_DZ_D$ and SM + SM $\to Z_D$ + SM processes is shown in dashed green.  Note that the dashed blue line connects the early and late LIDM regions.}
\label{fig:epsvsalphasols}
\end{center}
\end{figure}
 
In Fig.~\ref{fig:epsvsalphasols}, the possible solutions that provide the correct relic abundance in $\alpha_D$ vs $\epsilon$ parameter space are shown for three different choices of $m_\chi$.  At high mass, an appropriate choice of $\alpha_D$ and $\epsilon$ could realize any one of the solutions.  At moderate masses of $m_\chi\lesssim 3$ TeV, there is no longer a dominantly freezein solution (unless there was never a formation of the dark vector plasma).  For smaller masses $m_\chi\lesssim 10$ GeV, there is no valid reannihilation solution because the dark matter abundance produced through freezein processes is either too small to account for the dark matter density or injected into a dark vector plasma where the dark matter is still in equilibrium.  In this case, the early leak-in solution smoothly joins the late  leak-in solution.  

Generically, there is only one viable solution for $\alpha_D$ for a given $m_\chi$ and $\epsilon$.  However, at very high mass there is a region where different values of $\alpha_D$ can provide a mostly late LIDM, mostly freezein, or reannihilation solution; these multiple solutions are manifest in Fig.~\ref{fig:epsvsalphasols} where the curve for $m_\chi = 100$ TeV becomes non-monotonic.  In the sliver of parameter space where all three solutions are valid, $\alpha_{reann} \gg \alpha_{FI}\gg \alpha_{LI}$.  This three-solution region is the only place in the parameter space that a mostly freezein solution can be found.   However, in this construction we have implicitly assumed that the reheat temperature of the universe is large enough that $\tilde T_{RH}$, as dictated by the attractor solution, is larger than the dark matter freezeout temperature.  If the reheat temperature was too low, or the hidden sector did not internally thermalize, freezein solutions can occur.
%

\section{Signals of $B-L$ vector portal LIDM}
\label{sec:signals}

Despite the small size of the $B-L$ portal coupling $\epsilon$, there are many experimental handles on vector portal leak-in dark matter. In this section, we will discuss current limits on and potential future sensitivities to this parameter space.

\subsection{Indirect detection}
\label{sec:id}

The same process that allows LIDM to freezeout can facilitate dark matter annihilation throughout the universe's history, including today.  Indirectly detecting dark matter through these annihilation products is one of the most promising ways to probe LIDM models as the annihilation cross-section, which is $s$-wave in the $B-L$ vector portal model, does not depend directly on the very small coupling to the SM particles $\epsilon$.   

Additionally, the exchange of light mediators can enhance the tree-level annihilation cross-section from Eq.~\ref{eq:DMann} via the Sommerfeld effect \cite{Sommerfeld:1931,Hisano:2002fk,Hisano:2003ec,Hisano:2004ds}. 
 The $s$-wave cross-section can be expressed as \cite{Cirelli:2007xd,ArkaniHamed:2008qn,Tulin:2013teo,Evans:2017kti}
\beq
\vev{\sigma v} =  \vev{ S(\alpha_D, r,v) \sigma_{\chi\bar\chi\to Z_DZ_D} v} \approx S(\alpha_D, r,v_c) \vev{\sigma_{\chi\bar\chi\to Z_DZ_D} v},
\eeq
where $r = m_{Z_D}/m_{\chi}$, $v_c$ is some characteristic dark matter velocity for the system of interest, and the Sommerfeld enhancement factor for a Hulth\'en potential (a good approximation to a Yukawa potential with nicer analytic properties \cite{Cassel:2009wt}) is \cite{Evans:2017kti}
 \beq  
S(\alpha_D, r,v) = \frac{2\pi \alpha_D}{v} \frac{\sinh\left[\frac{6 v}{\pi r}\right]}{\cosh\left[\frac{6 v}{\pi r}\right] - \cosh\left[\sqrt{\frac{36 v^2}{\pi^2 r^2}-\frac{24 \alpha_D}{r}}\right]}.
\label{eq:SE}
\eeq
Low velocities and large couplings can give rise to sizable deviations from the non-Sommerfeld enhanced, tree-level cross-section.  DM annihilation products can produce signals in the Alpha Magnetic Spectrometer (AMS-02), Fermi Large Area Telescope (Fermi-LAT), or (indirectly) in various experiments that have measured the power spectrum of the Cosmic Microwave Background (CMB).   Following Ref.~\cite{Evans:2017kti}, we will use $v_{CMB}=10^{-7}$, $v_{dwarf}=10^{-4}$, and $v_{MW}=1.7\times 10^{-3}$ for the characteristic velocities of these systems in the Sommerfeld enhancement (Eq.~\ref{eq:SE}) to place constraints on the parameter space.  We do not consider the influence of the Sommerfeld effect on freezeout, as this would primarily affect only the large $\alpha_D$ region which corresponds not to leak-in, but reannihilation, discussed in Sec.~\ref{sec:others}, that produces the bulk of the relic abundance. 

Some of the most stringent constraints on annihilating dark matter come from the detailed measurements of the CMB power spectrum \cite{Ade:2015xua,Slatyer:2015jla}.  Injection of energetic charged particles and photons into the plasma can distort the CMB anisotropies.
  Planck, SPT, ACT, and WMAP results restrict the power injected into the CMB from DM annihilation, per DM mass, to satisfy $f_{eff}(m_\chi) \vev{\sigma  v}/{m_\chi} < 14$ pb c / TeV \cite{Slatyer:2015jla}, which allows for robust bounds to be placed on dark matter models.  The effective energy deposition efficiency $f_{eff}(m_\chi)$ \cite{Slatyer:2015jla, Madhavacheril:2013cna} depends on the branching fractions into specific annihilation channels, but it is $0.4-0.6$ for electron- and photon-enriched annihilations, small for neutrinos, and typically $\sim 0.2$ for everything else in the SM.  Despite the smallness of the energy deposition efficiency for neutrinos, at high DM masses neutrino-induced energy deposition into the CMB can be large enough that even dark vectors that are only able to decay to neutrinos are excluded.

\begin{figure}
\begin{center}
\includegraphics[width=0.48\textwidth]{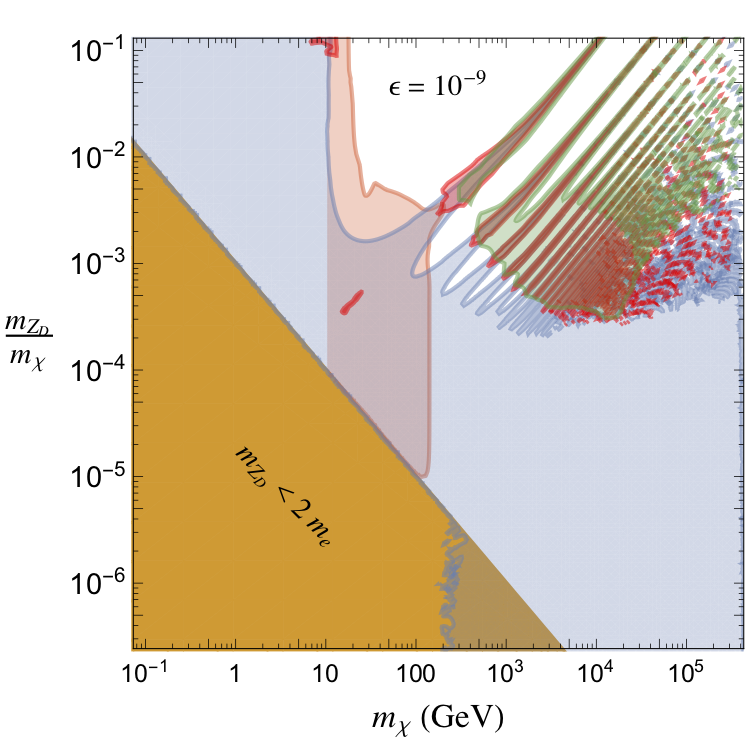} \hspace{1mm}
\includegraphics[width=0.48\textwidth]{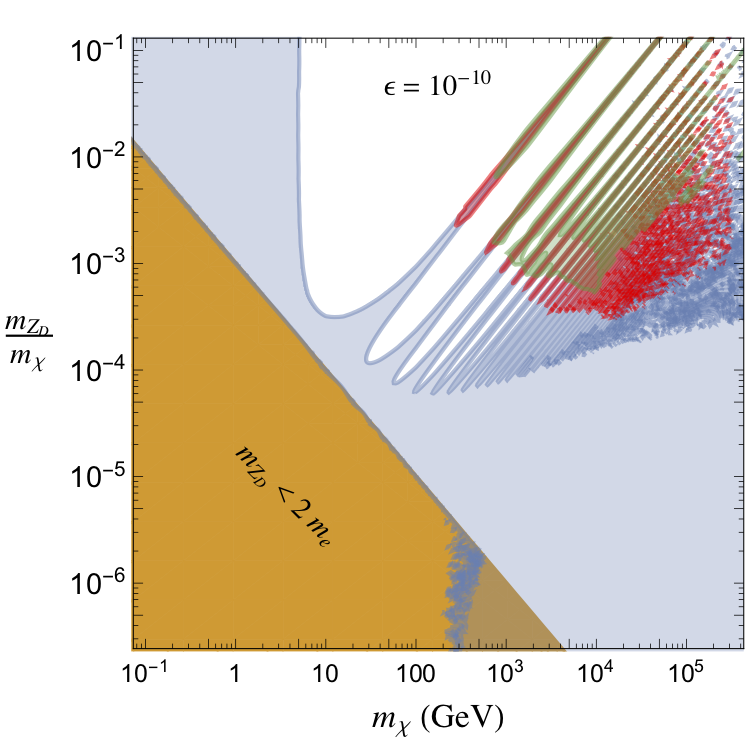}
\includegraphics[width=0.48\textwidth]{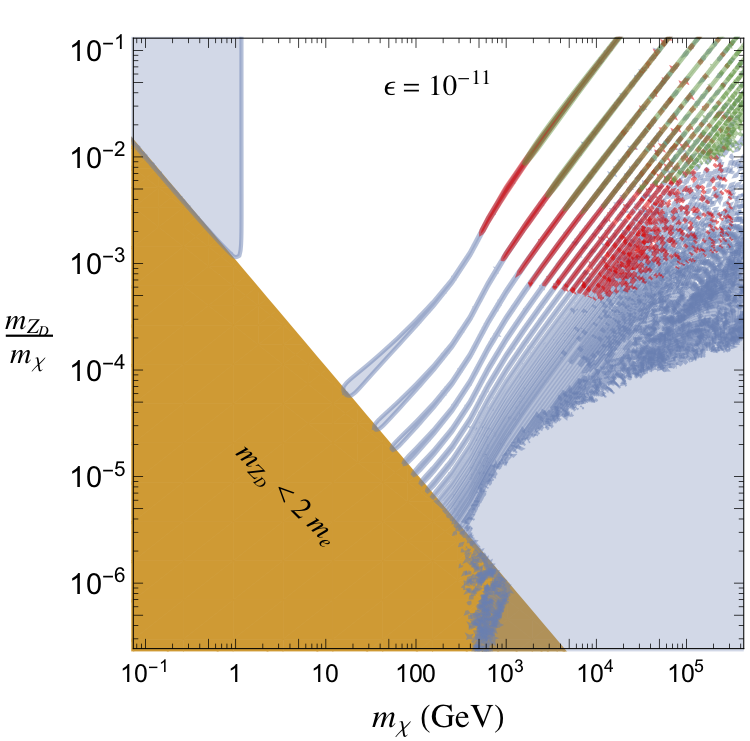} \hspace{1mm}
\includegraphics[width=0.48\textwidth]{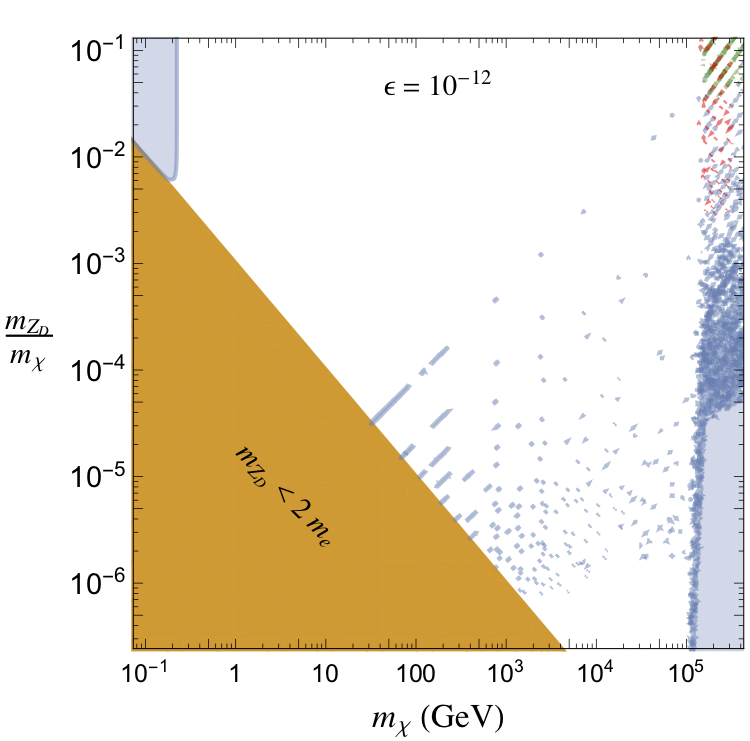}
\caption{Indirect detection constraints in the plane $m_{Z_D}/m_\chi$ vs $m_\chi$ for fixed $\epsilon$ ({\bf Upper Left:} $\epsilon=10^{-9}$; {\bf Upper Right:}  $\epsilon=10^{-10}$; {\bf Lower Left:}  $\epsilon=10^{-11}$; {\bf Lower Right:} $\epsilon=10^{-12}$).  Constraints described in the text are from the CMB (blue),  AMS-02 positrons (pink, upper left figure only), Fermi-LAT dwarfs (red),  H.E.S.S.\ galactic center (green). 
 The abrupt feature at 100 TeV on the lower right figure is when the model moves from a leak-in dark matter solution to a reannihilation solution and the annihilation cross-section jumps accordingly.
}
\label{fig:IDconstraints}
\end{center}
\end{figure}

Despite the current excess above predictions \cite{Aguilar:2013qda}, the observed positron flux at the AMS-02 experiment can be used to constrain dark matter annihilations that result in positrons.  We follow  \cite{Elor:2015bho}  in choosing to bound $\vev{\sigma v}\times \mathrm{Br}(Z_D\to e^+e^-)$, which is most stringent for dark vectors in the range  $2m_e<m_{Z_D}<2m_\mu$.

The Fermi Large Area Telescope experiment (Fermi-LAT) \cite{Atwood:2009ez} has observed gamma ray spectra for many dwarf galaxies \cite{Fermi-LAT:2016uux}, including many ultra-faint dwarf galaxies observed by the Dark Energy Survey (DES) \cite{Abbott:2005bi}.  As several dwarfs have low noise and large astrophysical $J$-factors, 
these observations can severely constrain dark matter annihilations \cite{Ackermann:2015zua}.  We use log-likelihood-ratios (LLR) provided by the Fermi collaboration for 24 energy bins for the 41 dwarf galaxies within the nominal sample of Ref.~\cite{Fermi-LAT:2016uux}.  We approximate the effect of correlated systematics by applying a 0.5$\sigma$ downward shift to the $J$-factors for each dwarf galaxy, as this was determined to closely replicate the limits placed by Fermi \cite{Evans:2017kti}.  The 41 dwarf LLRs are combined within each energy bin.  We generate the gamma ray spectra from dark matter annihilation in Pythia 8 \cite{Sjostrand:2014zea} at each point in $m_\chi$ vs $m_{Z_D}$.  All 24 bins are combined to form the $\chi^2$ with one degree of freedom to place limits on the annihilation rate $\vev{\sigma v}$.

Additionally, the observations of the galactic center performed with the High Energy Stereoscopic System (H.E.S.S.)\ experiment \cite{Bernloehr:2003vd} can place tighter constraints on heavier dark matter ($m_\chi \gtrsim1$ TeV) than Fermi due to the very large $J$-factors expected at the galactic center.\footnote{We use the Navarro-Frenk-White (NFW) $J$-factors, but if dark matter has a cored distribution, e.g., a Burkert profile \cite{Burkert:1995yz}, as some observations suggest may be the case \cite{BlaisOuellette:2000ma,Borriello:2000rv,deBlok:2001rgg,Swaters:2002rx,Gentile:2004tb,Gentile:2005de,Oh:2010mc,Rodrigues:2017vto}, then the limits obtained from H.E.S.S.\ would be unrealistically stringent.}   As H.E.S.S.\ does not provide their data, we follow Refs.~\cite{Profumo:2016idl,Profumo:2017obk} and use the 112 hour data from a gamma ray study \cite{Abramowski:2011hc} to obtain an observed gamma ray spectra for the signal and background regions and simply scale these results up to 254 hours of data to project fairly conservatively what an updated study could achieve.  We again use Pythia 8 \cite{Sjostrand:2014zea} to generate the annihilation signal gamma ray spectra at each point in $m_\chi$ vs $m_{Z_D}$.  The effective area was collected from \cite{HESSeffareaTalk}.  With the statistical procedure outlined in \cite{Lefranc:2015vza}, we derive a $\chi^2$ with one degree of freedom to approximate the limits on the annihilation rate $\vev{\sigma v}$ that H.E.S.S.\ would be able to find.  While this procedure allows us to place approximate limits on the model, firmer statements would be possible if H.E.S.S.\ were to provide the tools required to reliably recast their results, e.g.\ by providing the LLRs for a signal + background hypothesis for each energy bin as a function of injected signal strength.

The interplay of these constraints  in the parameter space is illustrated in Fig.~\ref{fig:IDconstraints}.  For four fixed values of $\epsilon$ (at $10^{-9}$, $10^{-10}$, $10^{-11}$, and $10^{-12}$), we show CMB constraints (blue), positrons from AMS-02 (pink), gamma-rays from dwarf galaxies at Fermi-LAT (red), and gamma rays from the galactic center at H.E.S.S.\ (green).  In the lower left corner of each figure, the dark vector is below $2m_e$ and decays to neutrinos. While most indirect detection constraints considered here are insensitive to neutrinos, the CMB power spectrum can be sufficiently distorted by very energetic neutrinos that arise from heavy dark matter annihilations \cite{Slatyer:2015jla}.    The sharp transition near 100 TeV in the lower right panel occurs where  the model moves from leak-in dark matter to reannihilation and the cross-section jumps due to the larger $\alpha_D$ needed for the correct relic abundance.

Finally, Fermi-LAT observations of the smooth galactic halo may place more stringent constraints than dwarfs \cite{Chang:2018bpt}.  The Cherenkov Telescope Array (CTA) \cite{Consortium:2010bc} would greatly enhance the sensitivity to heavy dark matter \cite{Silverwood:2014yza}.  A full treatment of these (potential) limits is beyond the scope of this work.

\subsection{Direct detection}
\label{sec:dd}

 Despite the smallness of the $B-L$ portal coupling  $\epsilon$, direct detection experiments 
 can be an important probe of leak-in dark matter.  Dark vector exchange contributes to the 
 non-relativistic,  spin-averaged amplitude-squared for DM-nucleus scattering, which can be 
 written as  
\beq
\abs{\bar{ \mathcal M}^{NR}(E_R)}^2 = \abs{\frac{\mathcal M}{4m_\chi m_N}}^2 = 
 g_D^2 \epsilon^2 A^2 F^2(E_R)    \abs{\frac{1}{m_{Z_D}^2+2 m_N E_R} }^2 
 \label{eq:DDamp}
 \eeq 
where $E_R$ is the nuclear recoil energy,  $m_N$ and $A$  
are the mass and mass number, respectively, for the target nucleus (Xenon, in the 
case of interest), $F^2(E_R)$ is the nuclear form factor, for which we take the Helm form factor 
\cite{Helm:1956zz,Lewin:1995rx}.  When  
$m_{Z_D} \lesssim  v_\chi m_\chi m_N / (m_\chi +m_N)$,
the recoil energy dependence in the propagator 
is necessary to properly track the transition into a long-range interaction.

\begin{figure}
\begin{center}
\includegraphics[width=0.6\textwidth]{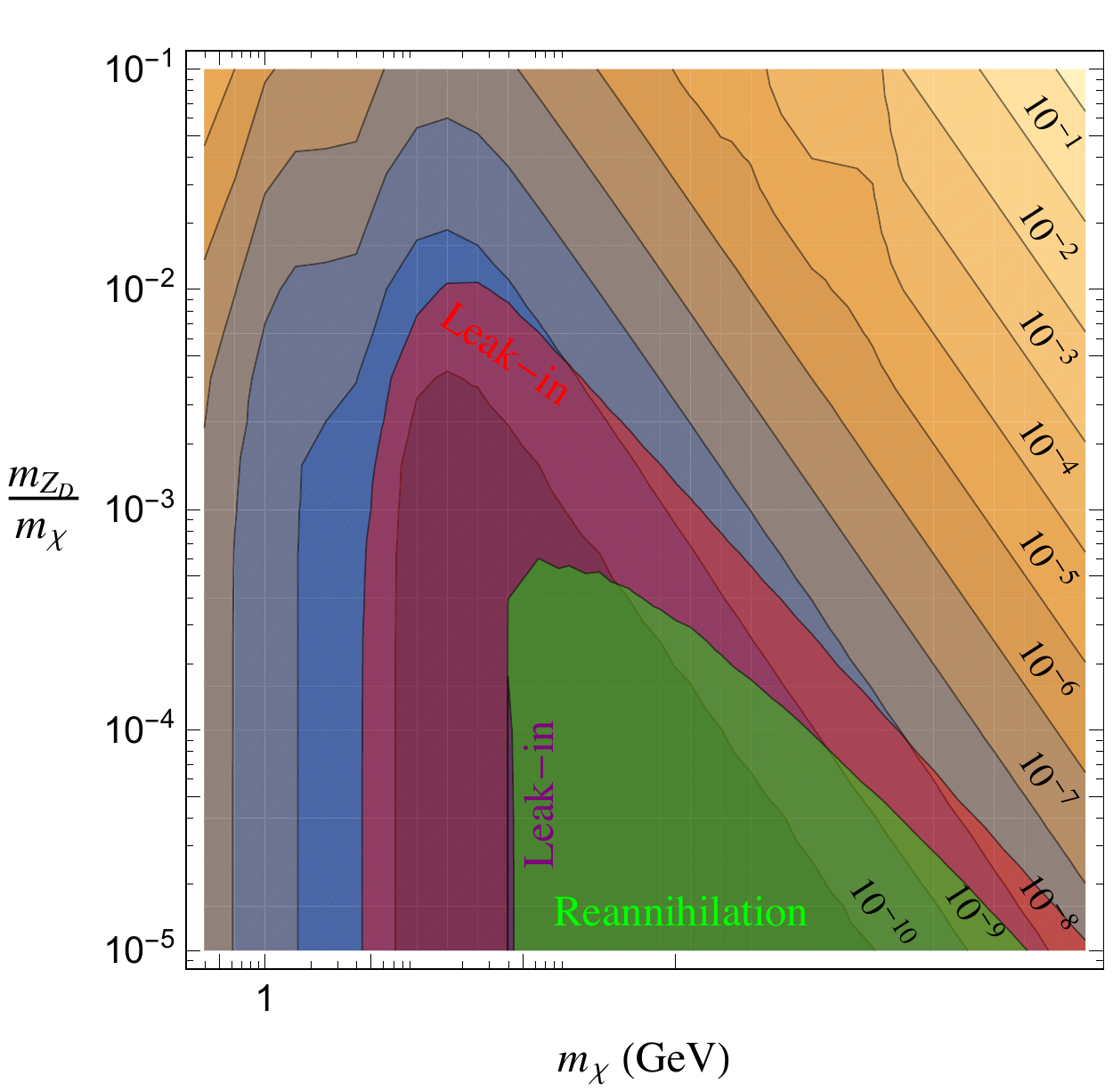}
\caption{Direct detection at XENON 1T, DarkSide-50 and CRESST constrain the portal coupling $\epsilon$ to be smaller than the values shown by the contours in the plane of $r=m_{Z_D}/m_\chi$ vs $m_\chi$.  For $r\ll1$, the constraints force the model below the equilibration floor. In red, we show where the model is required to be at or below the late leak-in dark matter scenario.  In green, the model is required to be in reannihilation or below, while the narrow purple sliver is in early leak-in.  See Fig.~\ref{fig:montanaroadmap} for more details.}
\label{fig:DDconstraints}
\end{center}
\end{figure}

When the amplitude is independent of the DM velocity, 
the event detection rate of the experiment per unit detector mass can
be written as \cite{Fan:2010gt,Freese:2012xd}
\beq
R\lp\bar{\mathcal M}^{NR}(E_R)\rp = \frac{\rho_\chi}{2\pi m_\chi} \int_{0}^\infty dE_R \abs{\bar{\mathcal M}^{NR}(E_R)}^2  \epsilon(E_R) \eta(E_R).
\label{eq:DDrate}
\eeq
where $\rho_\chi=0.3\, \text{GeV}/\text{cm}^3$ is the local DM density, $\epsilon(E_R)$ is the selection efficiency specific to the experiment, and the mean inverse speed $\eta$  is defined by \cite{Freese:2012xd}
\beq
\eta(E_R) = \int_{v>v_{min}(E_R)} \frac{f(v)}{v} d^3v
\eeq
for which we use the expression in Ref.~\cite{Lewin:1995rx} to match the experiments (and not the more accurate expression found in Ref.~\cite{Freese:2012xd}).  
If it is reasonable to approximate $\bar{\mathcal
  M}^{NR}(E_R)\to \bar{\mathcal M}^{NR}(0)$ in Eq.~\ref{eq:DDrate}
  (as is typical for contact interactions),
then the particle physics inputs may be factorized from the astrophysical 
and experimental inputs.  Most direct detection results are presented 
using a cross-section that has been both factorized in this manner and 
posed in terms of an effective cross-section per nucleon.  Defining the 
reduced mass of the nucleon-DM system as $\mu_{\chi n}=m_\chi m_n/(m_\chi + m_n)$, 
the per-nucleon-DM cross-section in this model is
\beq
\sigma_{\chi n}^0 = \frac{ \mu_{\chi n}^2 \abs{\bar{\mathcal M}^{NR}(0)}^2}{\pi A^2} =  \frac {4\alpha_D \epsilon^2 \mu_{\chi n}^2}{ m_{Z_D}^4}.
\label{eq:DDexclusion}
\eeq

As can be seen from Eq.~\ref{eq:DDamp}, the assumption of recoil energy independence 
 breaks down when  
$m_{Z_ D}^2 \lesssim 2 m_N E_R \sim \mu_{\chi N}^2 v_\chi^2$.
We will determine the excluded cross-section via
\beq
\sigma_{\chi n} = \sigma_{\chi n}^0 \frac{R\lp\bar{\mathcal M}^{NR}(E_R)\rp}{R\lp\bar{\mathcal M}^{NR}(0)\rp},
\label{eq:DDnet}
\eeq
in order to correctly account for this important effect at low mediator masses.
The latest XENON1T limits \cite{Aprile:2018dbl} 
place the tightest constraints in the parameter space.  
We show the current limits and regions where direct 
detection forces the model below the equilibration floor 
in Fig.~\ref{fig:DDconstraints}.
Recent limits from DarkSide-50, CRESST-III, and EDELWEISS 
\cite{Agnes:2018ves,Abdelhameed:2019hmk,Armengaud:2019kfj} 
probe lighter masses, but are currently not sensitive enough to 
place meaningful constraints below the equilibration floor.
The sensitivity scales as $1/m_{Z_D}^4$, but saturates when
$m_{Z_D}^2 \lesssim 2 m_N E_{min} = (30\text{ MeV})^2$.  
Interestingly for dark matter above $100$ GeV, the limits from direct 
detection are nearly independent of the dark matter mass, and set 
the same constraint across the $m_{Z_D}$ vs $\epsilon$ plane.  
This is because the dark matter flux drops as $1/m_{\chi}$, while 
$\sigma^0_{\chi n} \propto \alpha_D$, which for reannihilation also scales roughly as $m_\chi$.  

Several recent experiments, notably SENSEI \cite{Abramoff:2019dfb}, DAMIC \cite{Aguilar-Arevalo:2019wdi}, XENON10 \cite{Essig:2017kqs}, SuperCDMS \cite{Agnese:2018col}, and DarkSide-50 \cite{Agnes:2018oej}, have constrained very light dark matter scattering off of electrons.  The relevant cross-section for this is simply  \cite{Emken:2019tni}
\beq
\bar \sigma_{\chi e} =  \frac {4\alpha_D \epsilon^2 \mu_{\chi e}^2}{ m_{Z_D}^4},
\label{eq:DDexclusion4e}
\eeq
which for 1 MeV $\lesssim m_\chi \lesssim 1$ GeV is approximately
\beq
\bar \sigma_{\chi e} \sim 4 \text{ ab } \lp \frac{\alpha_D}{10^{-4}}\rp \lp\frac{\epsilon}{10^{-10}}\rp^2 \lp\frac{100 \text{ keV}}{m_{Z_D}}\rp^4.
\label{eq:DDexclusion4e}
\eeq
As we will uncover in the next two sections, $\epsilon\sim10^{-10}$ is the largest value possible and $\sim$100 keV is lightest that a $B-L$ vector can be for dark matter in this mass range.  With current constraints in the $\bar \sigma_{\chi e} \sim10$ fb range, there are several orders of magnitude further to probe before these electron recoil experiments could have sensitivity to this model, sensitivity which may be achievable by the proposed DAMIC-M \cite{Emken:2019tni}.

\subsection{Constraints on the $B-L$ vector boson}
\label{sec:dkphoton}

\begin{figure}
\begin{center}
\includegraphics[width=1\textwidth]{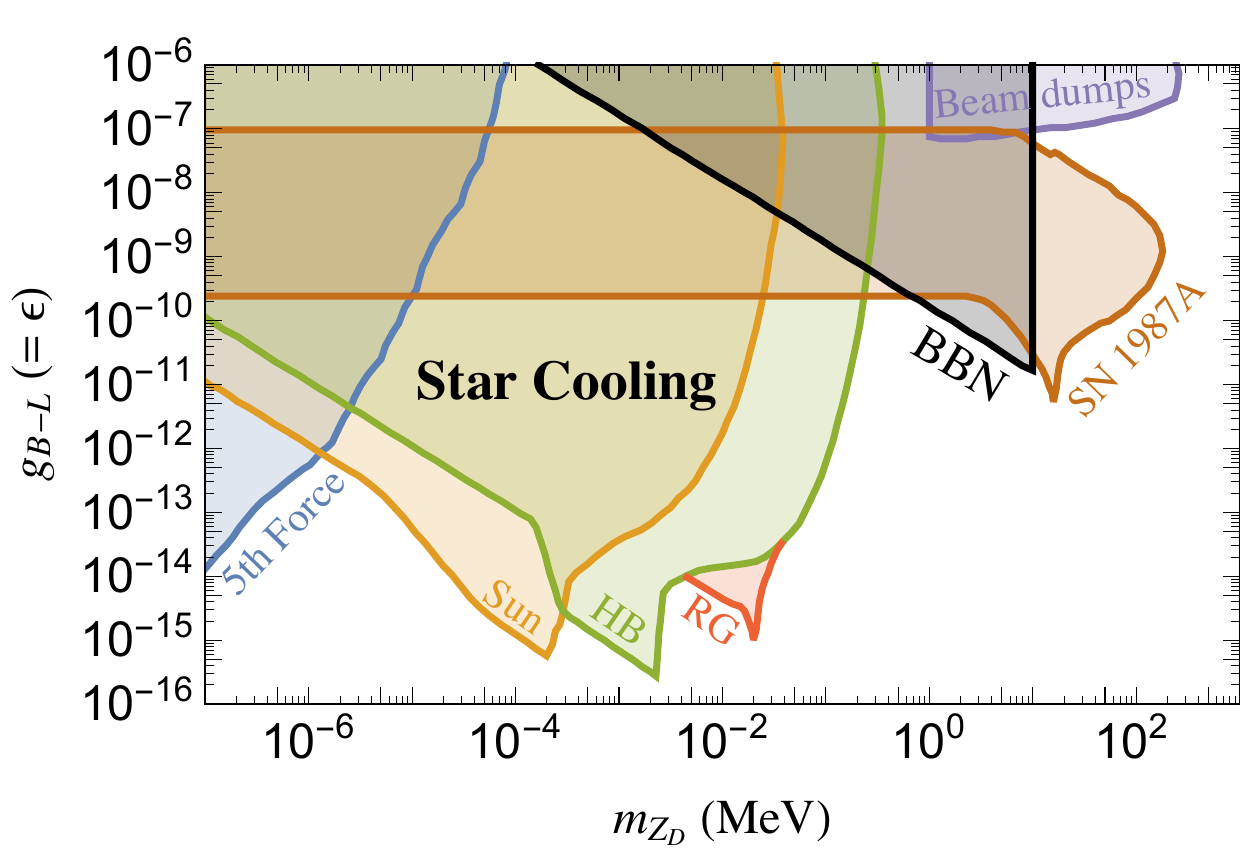}
\caption{Constraints on light, weakly coupled $B-L$ dark vectors. There are limits from (left to right) fifth force experiments \cite{Hardy:2016kme,Hoskins:1985tn,Kapner:2006si,Geraci:2008hb,Sushkov:2011zz}, stellar cooling through emission of dark vectors in the sun (Sun), horizontal branch stars (HB), and red giants (RG)  \cite{Hardy:2016kme,Knapen:2017xzo}, BBN \cite{Boehm:2013jpa,Knapen:2017xzo}, the cooling of supernova 1987A \cite{Chang:2016ntp,Knapen:2017xzo}, and beam dump experiments \cite{Bjorken:1988as,Athanassopoulos:1997er,Essig:2010gu,Andreas:2012mt}.}
\label{fig:BmLVectorConstraints}
\end{center}
\end{figure}

Independent of the nature of the dark matter, there is a wide variety of experimental searches for a $U(1)_{B-L}$ gauge boson.  In the region of dark vector masses and coupling strengths of interest for LIDM, $10^{-7} \gtrsim g_{B-L} \equiv \epsilon \gtrsim 10^{-14}$, the most important constraints come from fifth force experiments \cite{Hardy:2016kme,Hoskins:1985tn,Kapner:2006si,Geraci:2008hb,Sushkov:2011zz}, $N_{eff}$ constraints on the number of relativistic species present during BBN due to the dark $B-L$ vector maintaining thermal equilibrium with the neutrinos after decoupling from the electron-photon plasma, resulting in heating of the neutrino sector \cite{Boehm:2013jpa,Knapen:2017xzo}, the cooling of Supernova 1987A \cite{Chang:2016ntp,Knapen:2017xzo}, the electron beam dump E137 \cite{Bjorken:1988as,Andreas:2012mt}, the neutrino experiment LSND, interpreted as a proton beam dump \cite{Athanassopoulos:1997er,Essig:2010gu}, and  especially stellar cooling through emission of dark vectors in the sun (Sun), horizontal branch stars (HB), and red giants (RG)  \cite{Hardy:2016kme,Knapen:2017xzo}.  The current limits on weakly coupled $B-L$ vector bosons are summarized in Fig.~\ref{fig:BmLVectorConstraints}.

It is possible there are additional constraints both from SN1987A, where a $B-L$ dark vector decays to positrons that at late times contribute to 511 keV gamma ray signal \cite{DeRocco:2019njg}, and BBN, where $B-L$ dark vectors that are produced in the early universe, survive until BBN, then decay causing photo-disintegration of nuclei \cite{Fradette:2014sza}.  Derivation of the specific constraints for a $B-L$ dark vector is beyond the scope of this work.

\subsection{Dark matter self-interactions}
\label{sec:sidm}

$B-L$ vector portal LIDM can have sizable self-interactions, especially in the regime $r = m_{Z_D}/m_\chi \ll 1$.  
The most stringent limits on DM self-interactions in this model come from the measured ellipticity of galaxy halos, or (less precisely) through the generation of very large cross-sections on dwarf scales, far in excess of those cross-sections that yield acceptable dwarf galaxy properties in simulations.

Astrophysical observables are generally sensitive to 
 the transfer cross-section for $\chi\bar\chi \to \chi\bar\chi$ or $\chi\chi\to\chi\chi$ scattering,
\begin{align}
\sigma_{\text{T}} \equiv \int d\Omega \frac{d\sigma}{d\Omega}(1-\cos\theta),
\end{align}
where $\theta$ is the scattering angle in the CM frame.
In the $B-L$ vector portal model, the DM elastic scattering process is mediated by dark vector exchange. In the Born regime, which occurs when $\alpha_D\ll r$, there is an  analytic closed-form expression for the transfer cross-section valid for both $\chi\bar\chi \to \chi\bar\chi$ and $\chi\chi\to\chi\chi$ (calculated with Package-X \cite{Patel:2016fam}),
\begin{align}
\sigma_{\text{T}} =&\frac{\pi \alpha_{D}^2}{m_{\chi}^2} \left(4\frac{r^4 +4(1+2v^2)^2 +2r^2(2+3v^2)}{r^4+4r^2v^2} \right. \nonumber \\
&\hspace{12mm} \left.+\frac{2(1-4v^2-8v^4)-2r^2(2+3v^2)-r^4}{v^2(r^2+2v^2)}\ln\left(1+\frac{4v^2}{r^2}\right) \right).
\end{align}
In the classical Rutherford regime, the scattering is long-range, so that the momentum transfer is large compared to the mediator mass $v\gg r$, and non-perturbative since $\alpha_D \gtrsim r$.  In terms of $\beta\equiv 2\alpha_D r v^2 $, the transfer cross-section can be approximated as \cite{Cyr-Racine:2015ihg}
\begin{align}
 \sigma^{+}_{\text{T}}=
 \left\{
 \begin{array}{ll}
        \frac{2\pi}{m^2_{Z_D}}\beta^2\ln (1+\beta^{-2}), \quad \beta \lesssim 10^{-2}    \\
        \frac{8\pi}{m^2_{Z_D}}\frac{\beta^{1.8}}{1+5\beta^{0.9}+0.85\beta^{1.6}}, \quad 10^{-2} \lesssim \beta \lesssim 10^4 \\
        \frac{\pi}{m^2_{Z_D}}(\ln 2\beta -\ln\ln 2\beta)^2, \beta \gtrsim 10^4,
 \end{array}
\right.
\end{align}
for an attractive interaction and
\begin{align}
 \sigma^{-}_{\text{T}}=
 \left\{
 \begin{array}{ll}
        \frac{2\pi}{m^2_{Z_D}}\beta^2\ln (1+\beta^{-2}), \quad \beta \lesssim 10^{-2}    \\
        \frac{7\pi}{m^2_{Z_D}}\frac{\beta^{1.8}+280(\beta/10)^{10.3}}{1+1.4\beta+0.006\beta^{4}+160(\beta/10)^{10}}, \quad 10^{-2} \lesssim \beta \lesssim 10^2 \\
        \frac{0.81\pi}{m^2_{Z_D}}(1+\ln \beta -(2\ln\beta)^{-1})^2, \beta \gtrsim 10^2,
 \end{array}
\right.
\end{align}
for a repulsive interaction.  Since the symmetric DM in this model is composed of an equal number of particles and antiparticles, we will take $\sigma_{\text{T}}= \frac{1}{2}(\sigma_{\text{T}}^{+}+\sigma_{\text{T}}^{-})$. 

Between the Born regime and the classical regime is the resonant regime, characterized by $\alpha_D\gtrsim r$ and $v\sim r$, where the transfer cross-section has a complicated velocity dependence.  The transfer cross-section here can be calculated by summing up contributions from a sufficiently large number of partial waves \cite{Tulin:2013teo}. This procedure is computationally expensive, however, and in much of the parameter space we will be able to bypass computations in the resonant regime by employing a bounding method, described below.

\paragraph{Self-interaction cross-section on dwarf scales.}  
If dark matter has too large of a transfer cross-section, then the galactic properties produced in $N$-body simulations do not match observations.  Comparison of $N$-body simulations with observations suggest upper bounds on $\langle\sigma_T\rangle/m_\chi$ in dwarf systems of order $10 \,\mathrm{cm}^2/\mathrm{g} \approx 20\, \mathrm{barn/GeV}$ or below, both in constant cross-section models \cite{Rocha:2012jg,Zavala:2012us} and long-range models \cite{Kaplinghat:2015aga}. 

Meanwhile simulations (albeit of constant self-interaction cross-sections) indicate that cross-sections in excess of $50 \,\mathrm{cm}^2/\mathrm{g} \approx 100$ barn/GeV begin to exhibit core collapse in dwarf galaxies \cite{Vogelsberger:2012ku, Elbert:2014bma}.

To evaluate  $\vev{\sigma_{\text{T}}}$, we construct two separate transfer cross-sections thermally averaged over a Maxwellian velocity distribution with $v_{RMS}=30$ km$/$s, one assuming the Born cross-section $\langle \sigma^{B}_{\text{T}}\rangle$, and the other assuming the classical Rutherford cross-section $\langle \sigma^{R}_{\text{T}}\rangle$. For a given value of $m_\chi$ and $\epsilon$ (and thus $\alpha_{D}$ from the relic abundance condition), we solve for the value of $m_{Z_D}$ that realizes $\langle \sigma^{X}_{\text{T}}\rangle = 50 \,\mathrm{cm}^2/\mathrm{g} \times m_\chi$ for both cases ($X=B,R$).  
After solving for this minimum allowed $m_{Z_D}$ in both regimes, we check that the solution is self-consistent (specifically, we require $\alpha_D/r <10^{-1}$ in the Born regime  and in the Rutherford regime both $\alpha_D/r >1$ and $v/r>100$). 
This procedure then produces a curve in the $(m_{Z_D}, \epsilon)$ plane indicating (for a given $m_\chi$) where the specified thermally averaged transfer cross-section is obtained.  In  Figs.~\ref{fig:mainparasp1} and~\ref{fig:mainparasp2} we show curves for both $10 \,\mathrm{cm}^2/\mathrm{g}$ and $50 \,\mathrm{cm}^2/\mathrm{g}$, indicating where transfer cross-sections begin to exceed the values where $N$-body simulations accord with observations.  Parameter points with larger self-interaction cross-sections on dwarf scales are disfavored.

However, when the thermal averaging is performed in the classical regime, this procedure is not completely accurate, as the classical expression for $\sigma_{\text{T}}$ is only valid when $ v \gg r$. For sufficiently small relative velocities, scattering occurs in the resonant regime instead. In order to overcome this issue, we consider the following bounding method which will also let us largely bypass the necessity of calculating the transfer cross-section in the resonant regime. Consider the thermally averaged cross-section in the non-perturbative regime:
\begin{align}
\langle\sigma_{\text{T}}\rangle = \int_0^{\infty}d^3v f(v) \sigma_{\text{T}} = \int_{100r}^{\infty}d^3v f(v) \sigma^{R}_{\text{T}} + \int_0^{100r}d^3v f(v) \sigma^{\text{resonant}}_{\text{T}},
\end{align}
where we split the integral into classical and resonant contributions. Since both terms are non-negative, we obtain the following lower bound:
\begin{align}
\int_{100 r}^{\infty}d^3v f(v) \sigma^{R}_{\text{T}} \leq \langle\sigma_{\text{T}}\rangle.
\end{align}
In order to construct an upper bound, we note from \cite{Cyr-Racine:2015ihg} that the classical cross-section is an overestimate in the resonant regime. Hence, $\sigma^{R}_{\text{T}} \geq \sigma^{\text{resonant}}_{\text{T}}$, so that
\begin{align}
\langle\sigma_{\text{T}}\rangle &= \int_{100 r}^{\infty}d^3v f(v) \sigma^{R}_{\text{T}} + \int_0^{100r}d^3v f(v) \sigma^{\text{resonant}}_{\text{T}} \\
&\leq \int_{100r}^{\infty}d^3v f(v) \sigma^{R}_{\text{T}} + \int_0^{100r}d^3v f(v) \sigma^{\text{R}}_{\text{T}} = \int_{0}^{\infty}d^3v f(v) \sigma^{R}_{\text{T}}.
\end{align} 
As long as the scattering process is indeed non-perturbative, this method gives us a bounding region for the constraint curve.  The upper and lower bounds constructed in this manner often either closely coincide, or both lie deeply within excluded regions.  Only for DM masses around $m_\chi = 10$-$100$ GeV  do we need to explicitly evaluate the resonant contribution to the thermally-averaged transfer cross-section.  In  Figs.~\ref{fig:mainparasp1} and~\ref{fig:mainparasp2} we indicate with the blue hatched region the bound from resonant and/or classical (Rutherford) scattering, and with the red hatched region the bound from Born scattering.

\paragraph{Ellipticity.}  
DM self-interactions will tend to increase isotropy within galaxy haloes.  In particular, the measured ellipticity of the gravitational potential of the galaxy NGC720 \cite{Buote:2002wd} places a bound on DM self-interactions \cite{Feng:2009mn}.  
We here use a simple treatment of the ellipticity bound based on estimating the timescale $\tau_e$ for isotropizing the velocity dispersion in a halo and requiring that
it exceed the age of the universe \cite{Feng:2009mn},
\begin{align}
\label{eq:elliptimescale}
\tau_e = \frac{\langle E \rangle}{\langle\dot{E}\rangle}\geq 10^{10}\text{years}
\end{align}
where the average DM energy $E$ is given in terms of the velocity dispersion $v_0^2$ by
\begin{align}
\langle E \rangle = \frac{1}{2}m_{\chi}\langle v^2 \rangle = \frac{1}{2}m_{\chi} \frac{3}{2}v_0^2
\end{align}
(we take the velocity distribution to be locally given by a Maxwell-Boltzmann distribution).
Meanwhile the average energy transferred in a DM-DM collision is given by
\begin{align}
\label{eq:dotE}
\langle\dot{E}\rangle = \rho_{\chi}\int d^3\vec{v}f(v) v^3\sigma_{\text{T}}.
\end{align}
For simplicity we evaluate Eq.~\ref{eq:elliptimescale} with $\rho = 2.1$ GeV/cm$^3$ and $v_0=260$ km/s, 
corresponding to the middle of the range of values reported in Ref.~\cite{Feng:2009mn}.

For this model, the integral in Eq.~\ref{eq:dotE} is regulated by the finite dark vector mass. 
However, when the dark vector mass is sufficiently small compared to the momentum transfer, the integral will first be cut off by the net charge neutrality of the dark plasma, i.e., by requiring that the maximum impact parameter be smaller than the inter-particle spacing $\lambda_{pp}=(m_{\chi}/\rho_{\chi})^{1/3}$ \cite{Agrawal:2016quu}.  Thus $\tau_e$ becomes independent of $m_{Z_D}$ when
\beq
\frac{m_{Z_D}^2}{2m_{\chi}^2v_0^2 } \ll \frac{2}{1+y^2},
\eeq
where $y=\frac{m_{\chi}v_0^2}{\alpha_{\chi}}\lambda_{pp}$.

We use Eq.~\ref{eq:elliptimescale} as the constraint, which underestimates the time required to attain an isotropic distribution as it does not take into account the reduction in the energy transfer rate as initially anisotropic populations approach equilibration \cite{Agrawal:2016quu}. 
The resulting constraints are shown in brown in  Figs.~\ref{fig:mainparasp1} and~\ref{fig:mainparasp2}.
This bound is conservative for the purposes of identifying clearly allowed regions, but (as argued in Ref.~\cite{Agrawal:2016quu}) there are several ambiguities in translating the measured ellipticity of galaxy haloes  into bounds on DM self-interactions, making it hard to conclude that the shaded regions to the left of this bound are definitively excluded.

\subsection{Allowed parameter space}
\label{sec:results}

\begin{figure}[h]
\begin{subfigure}{0.5\textwidth}
  \hspace{-1.0cm}
  \includegraphics[width=1.0\linewidth]{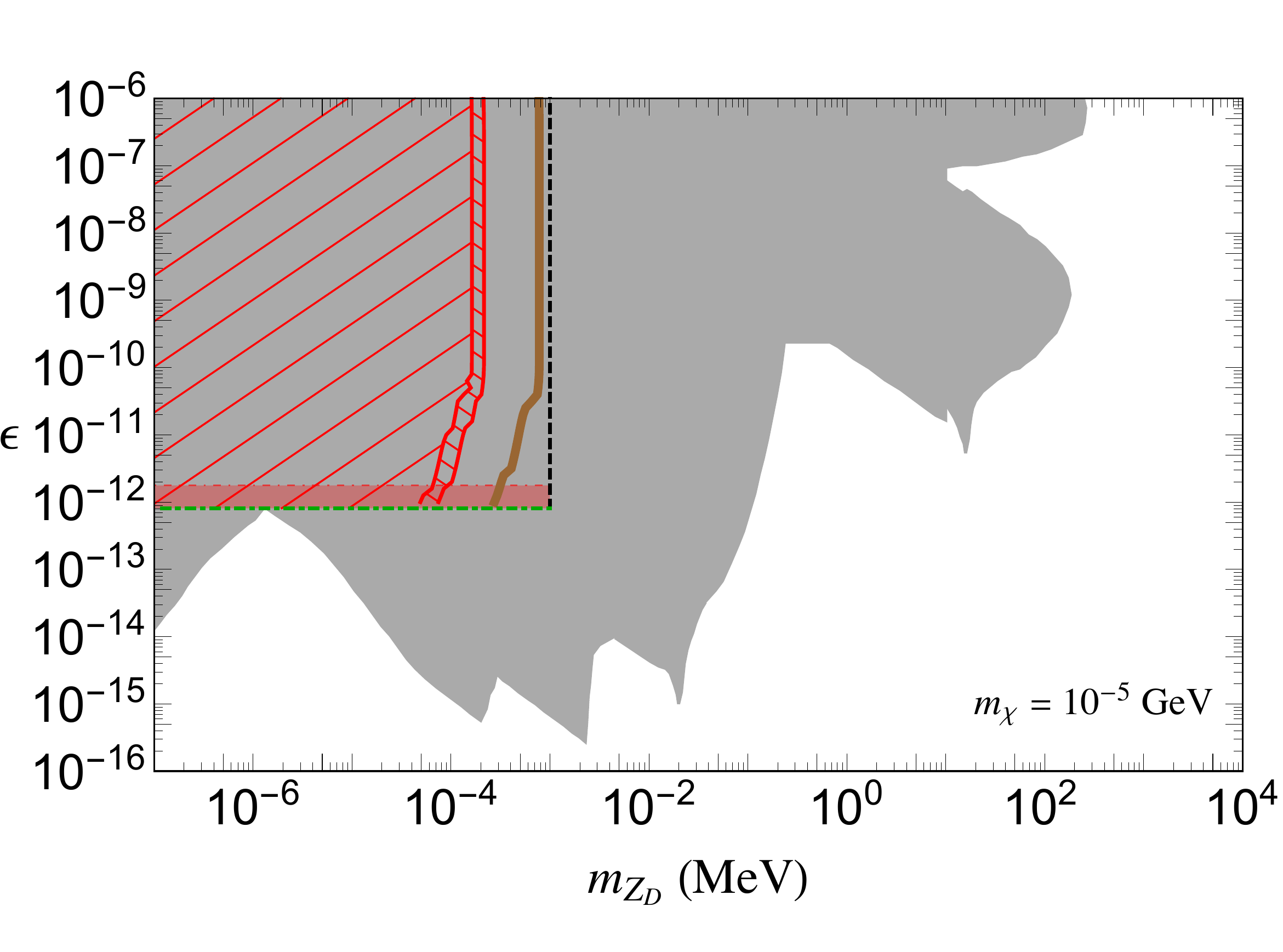}
\end{subfigure}%
\begin{subfigure}{0.5\textwidth}
  \hspace{-1.0cm}
  \includegraphics[width=1.0\linewidth]{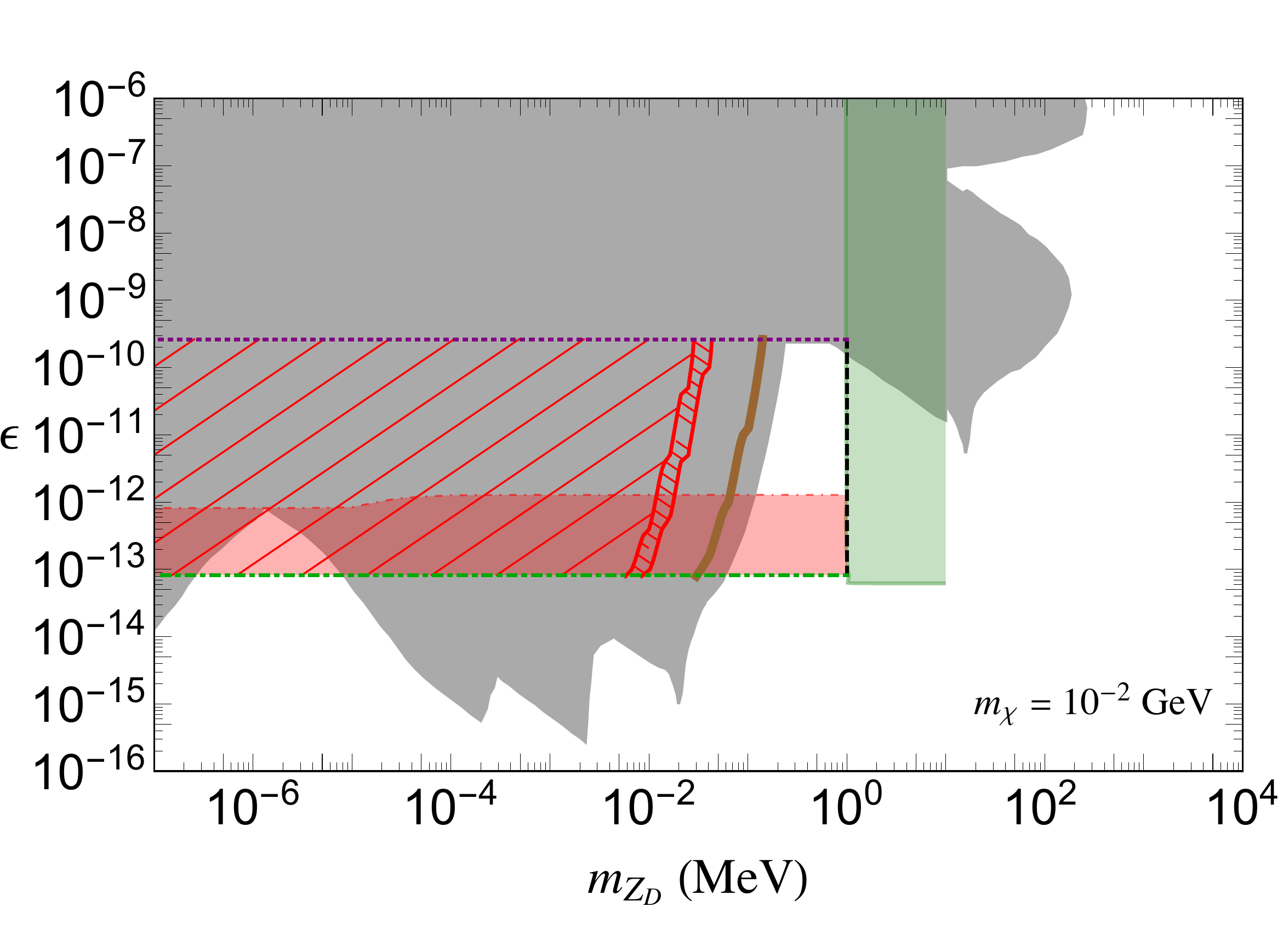}
\end{subfigure}
\begin{subfigure}{0.5\textwidth}
  \hspace{-1.0cm}
  \includegraphics[width=1.0\linewidth]{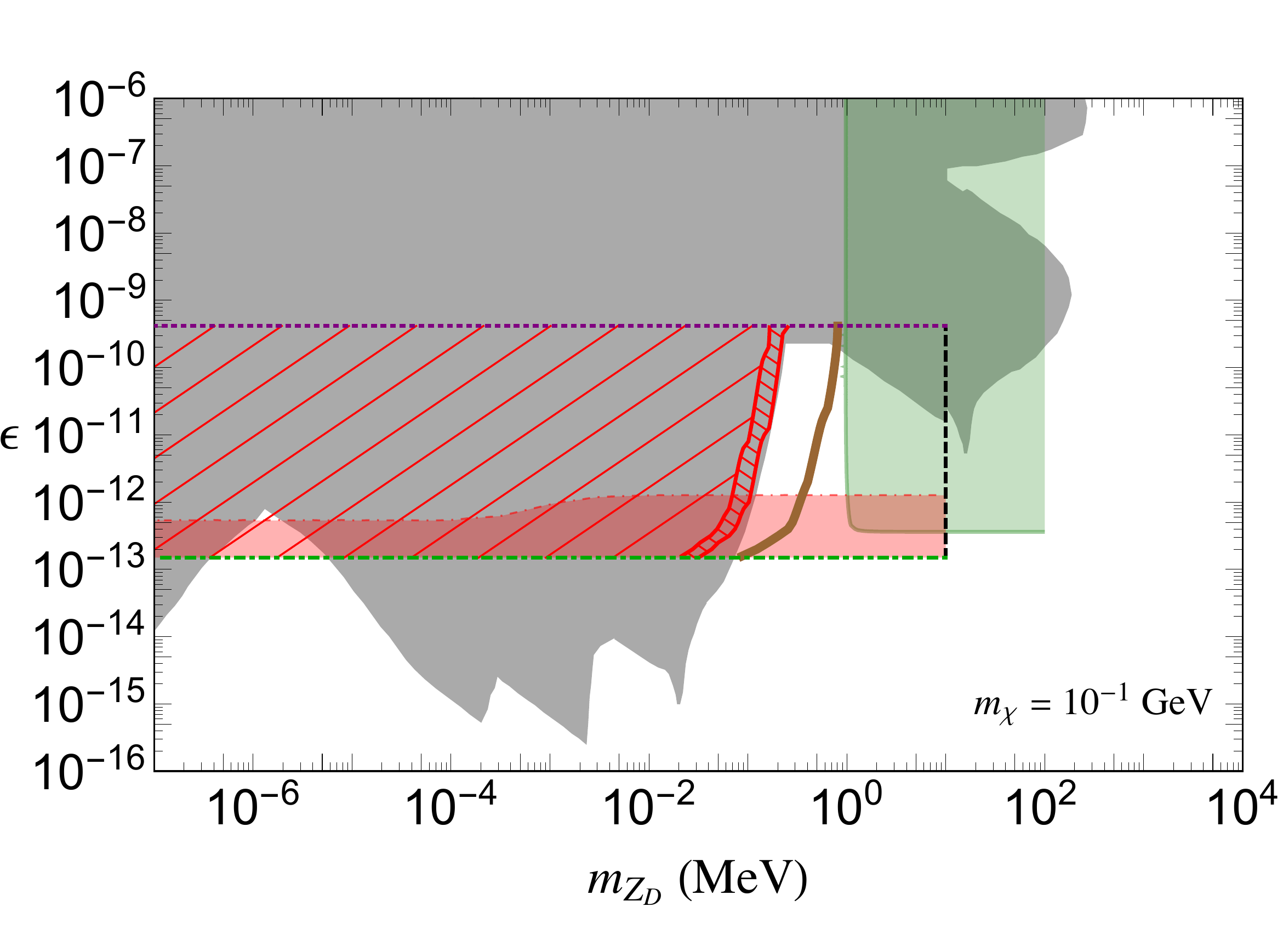}
\end{subfigure}
\begin{subfigure}{0.5\textwidth}
  \hspace{-1.0cm}
  \includegraphics[width=1.0\linewidth]{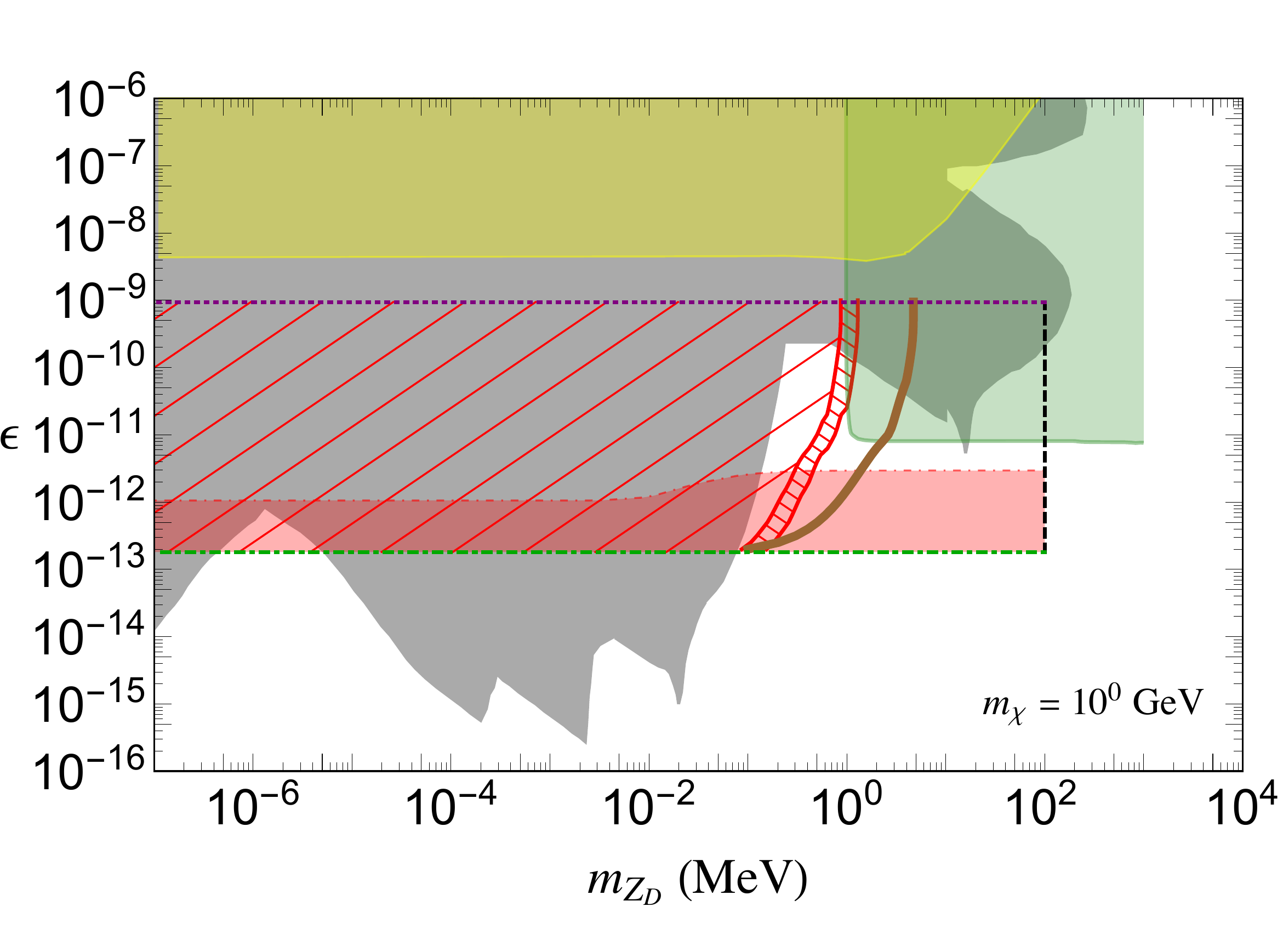}
\end{subfigure}
\caption{Observational constraints and surviving parameter space for $B-L$ leak-in dark matter in the $\epsilon$ versus $m_\zd$ plane, for fixed choices of $m_\chi$ ({\bf Upper Left:}  $m_\chi=10$ keV; {\bf Upper Right:}  $m_\chi=10$ MeV; {\bf Lower Left:}  $m_\chi=100$ MeV; {\bf Lower Right:}  $m_\chi=1$ GeV).  The dark shaded region shows the constraints on the $B-L$ boson  (Fig.~\ref{fig:BmLVectorConstraints}).  The horizontal purple dotted line shows the equilibration floor, the horizontal green dash-dotted line shows the absolute floor, and the vertical dashed black line indicates $m_\zd = 0.1 m_\chi$.  The light red region indicates regions that do not attain full internal thermal equilibrium.  
 Direct detection exclusions are shown in yellow (Fig.~\ref{fig:DDconstraints}), and CMB constraints in green  (Fig.~\ref{fig:IDconstraints}).  The red hatched region shows regions with $\langle \sigma_{\text{T}}\rangle/m_\chi > 50$ cm$^2/$g (left curve) and 10 cm$^2/$g (right curve), self-consistently computed in the Born regime (Sec.~\ref{sec:sidm}).  Regions to the left of the solid brown line violate the ellipticity bound on DM self-interactions (Sec.~\ref{sec:sidm}).
\label{fig:mainparasp1}
}
\end{figure}

\begin{center}
\begin{figure}[h]
\begin{subfigure}{0.5\textwidth}
  \hspace{-1.0cm}
  \includegraphics[width=1.0\linewidth]{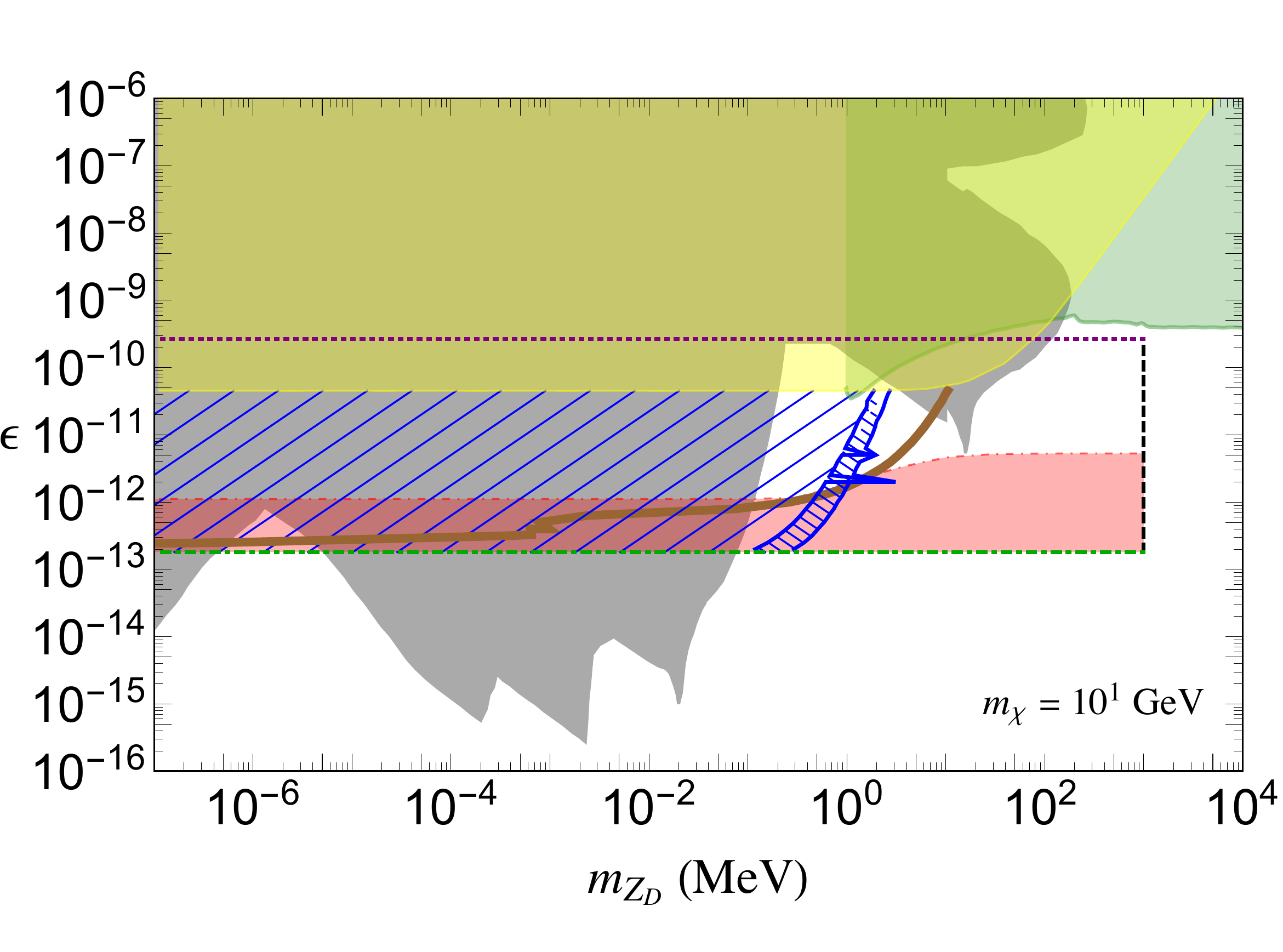}
\end{subfigure}
\begin{subfigure}{0.5\textwidth}
  \hspace{-1.0cm}
  \includegraphics[width=1.0\linewidth]{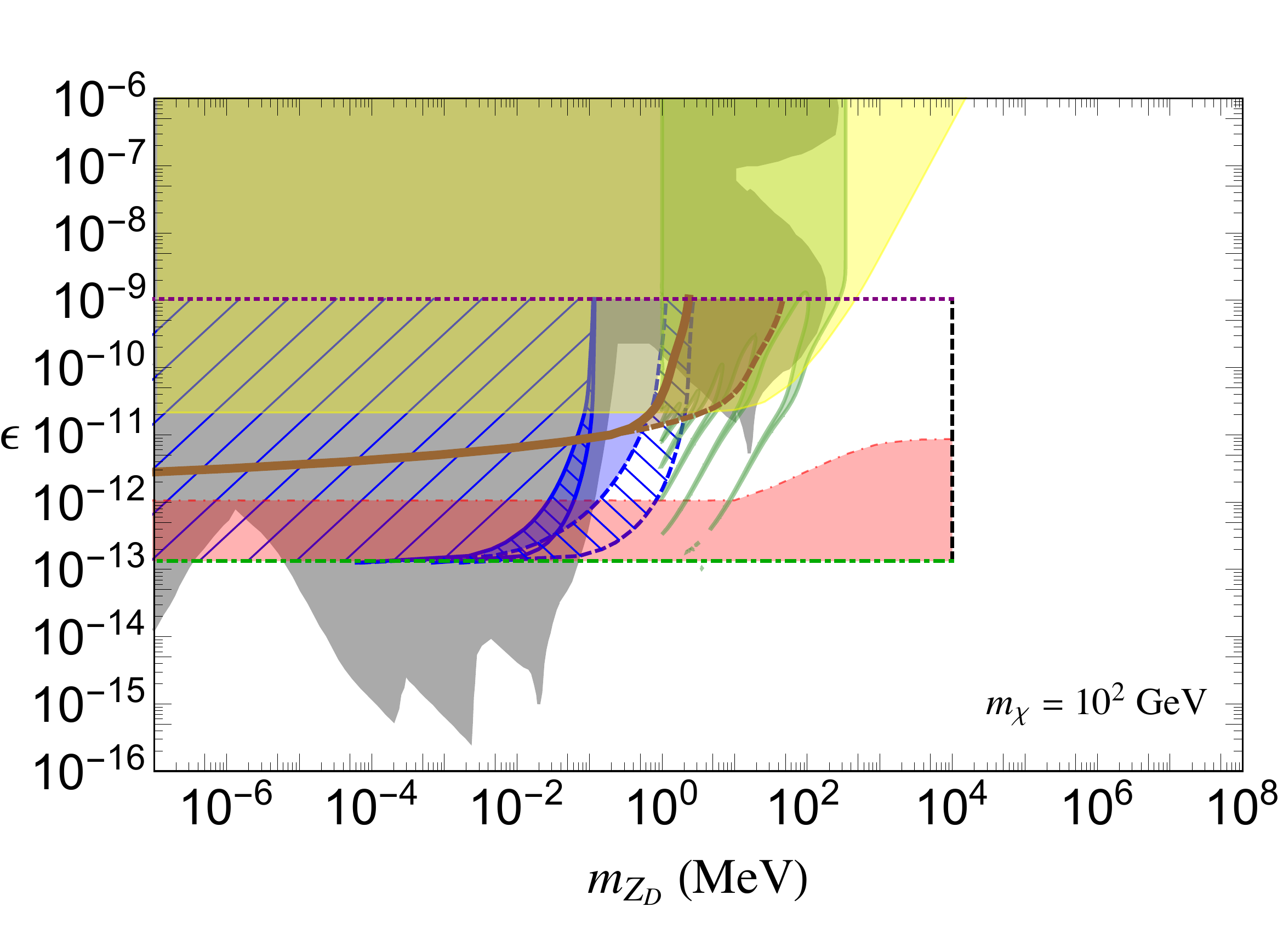}
\end{subfigure}
\begin{subfigure}{0.5\textwidth}
  \hspace{-1.0cm}
  \includegraphics[width=1.0\linewidth]{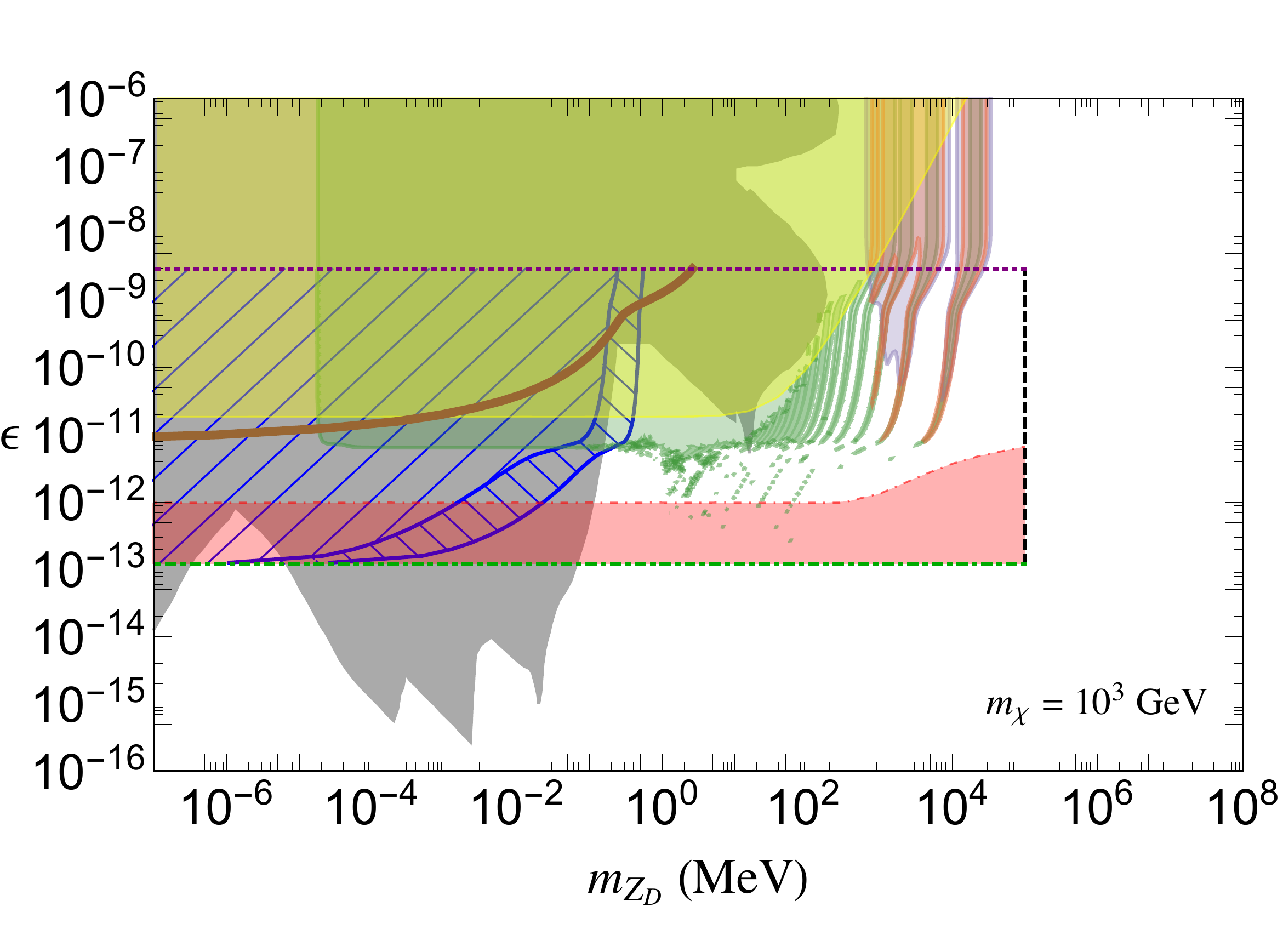}
\end{subfigure}
\begin{subfigure}{0.5\textwidth}
  \hspace{-1.0cm}
  \includegraphics[width=1.0\linewidth]{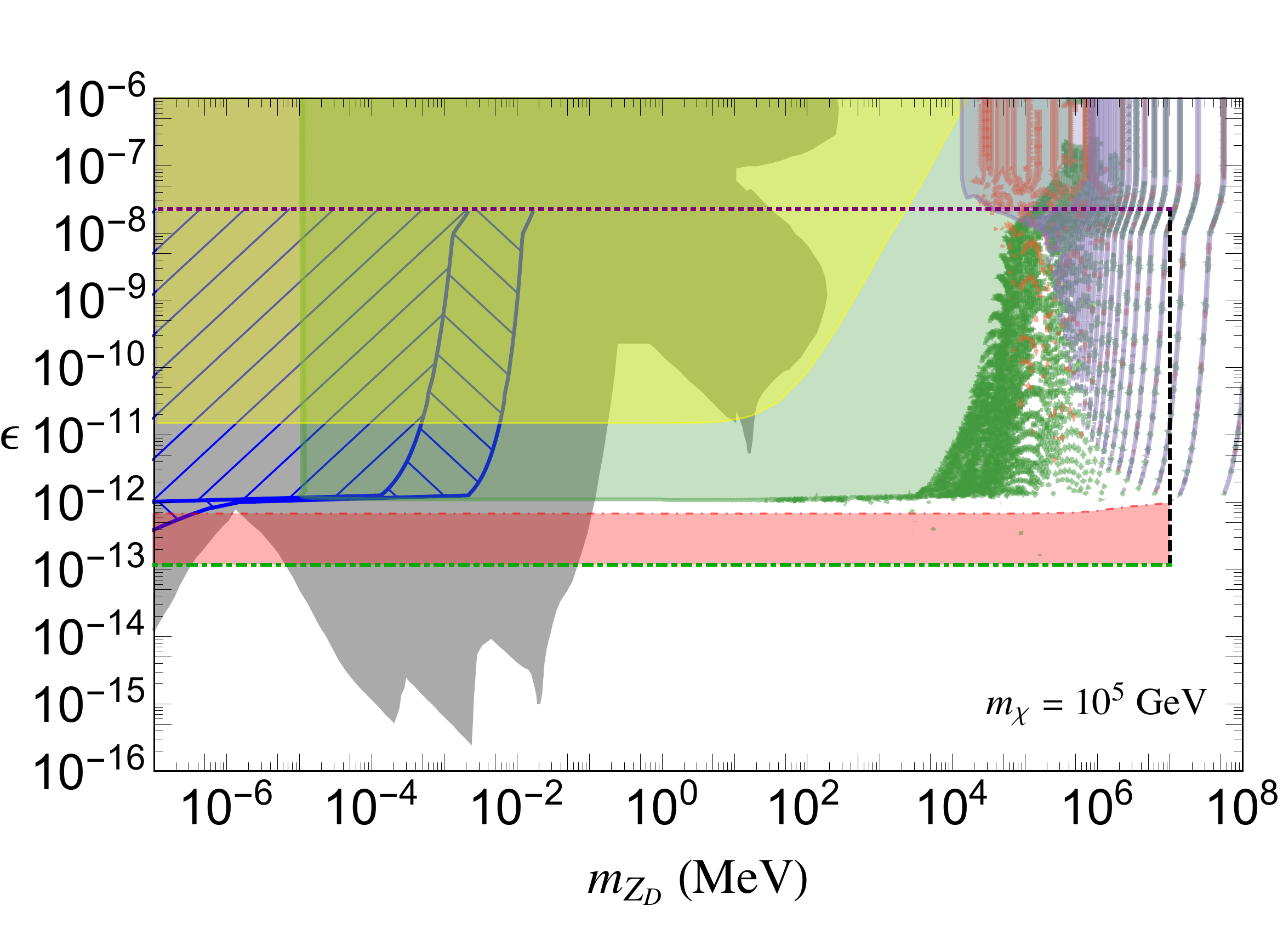}
\end{subfigure}
\caption{Observational constraints and surviving parameter space for $B-L$ leak-in dark matter in the $\epsilon$ versus $m_\zd$ plane, for fixed choices of $m_\chi$ ({\bf Upper Left:}  $m_\chi=10$ GeV; {\bf Upper Right:}  $m_\chi=100$ GeV; {\bf Lower Left:}  $m_\chi=1$ TeV; {\bf Lower Right:}  $m_\chi=100$ TeV).    Colors are as in Fig.~\ref{fig:mainparasp1}; in addition constraints from H.E.S.S. are shown in purple, and Fermi dwarfs in red.  The blue hatched region shows regions with $\langle \sigma_{\text{T}}\rangle/m_\chi > 50$ cm$^2/$g (left curve) and 10 cm$^2/$g (right curve), computed in the non-perturbative regime (Sec.~\ref{sec:sidm}).  In the upper right panel, for both dwarf and ellipticity constraints, solid lines denote known underestimates and dashed lines known overestimates; a full calculation in the resonant regime would yield a constraint in the shaded region in between. 
\label{fig:mainparasp2}
}
\end{figure}
\end{center}

The regions of dark vector parameter space consistent with LIDM  are shown for several different values of DM mass in Figs.~\ref{fig:mainparasp1} and~\ref{fig:mainparasp2}.  
For a fixed DM mass $m_\chi$, there is a specific region in the $(m_{Z_D},\epsilon)$ plane consistent with the LIDM mechanism.  For sufficiently large $\epsilon$, the SM and the HS attain thermal equilibrium before DM freezeout, while for sufficiently small $\epsilon$, DM will never obtain a a large enough co-moving number density to account for the relic abundance observed today.   Internal thermalization (see Appendix~\ref{sec:inttherm}) provides a more stringent, but less robust, condition than under-abundance; we will show lower bounds from both thermalization and absolute abundance  on the plots below. Meanwhile, the upper bound on $m_{Z_D}$ simply reflects the requirement that $r =m_{Z_D}/m_\chi \leq 0.1$, so that  (in the minimal model) the dark vector constitutes a relativistic radiation bath at DM freezeout.  

From Figs.~\ref{fig:mainparasp1} and~\ref{fig:mainparasp2}, we identify two distinct regions of parameter space consistent with dark vector constraints. First is an ``invisible'' region where the dark vector mass lies in the narrow window between stellar cooling bounds and CMB constraints on DM annihilations, $100 \,\mathrm{keV}\lesssim m_{Z_D} < 2m_e$.  In this regime, the dark vector decays entirely to neutrinos, rendering DM annihilation (largely) invisible to cosmic ray searches.  The second, ``visible'', region of parameter space occurs where $m_{Z_D}>2m_e$, and DM annihilation produces visible cosmic ray signals.  Stringent constraints on very light dark $B-L$ gauge bosons, combined with the excessively large DM self-interactions generated when $m_{Z_D}\lll m_\chi$, disfavor values of $m_{Z_D}$ below tens of keV.

For $m_\chi \lesssim 100$ MeV (Fig.~\ref{fig:mainparasp1}), the  combination of dark vector bounds, the restriction $m_{Z_D} \leq m_\chi/10$, the requirement of internal thermalization (light red), and CMB constraints on DM annihilation (green)  leave only the small invisible region available.  The narrow window of surviving parameter space can realize DM self-interaction cross-sections large enough to violate the ellipticity bound (brown line), further limiting the portions of parameter space that are clearly viable. For $m_\chi = 1$ GeV, a narrow strip of parameter space with visible DM annihilations opens up between the  CMB and internal thermalization constraints.

For heavier DM,  $m_\chi \gtrsim$ GeV (Fig.~\ref{fig:mainparasp2}), both visible and invisible regions are allowed.  For $m_\chi = 10$ and 100 GeV, direct detection experiments are the only probe of the invisible region of parameter space, where indirect detection searches have no reach.  
In the top left plot, for  $m_\chi = 10$ GeV, we perform a full resonant calculation for both $\langle \sigma_{\text{T}}\rangle/m_\chi$ and ellipticity.  
In the top right plot, for  $m_\chi = 100$ GeV, the left (right) solid blue line and left (right) dashed blue line
form the brackets for the classical regime calculation saturating $\langle \sigma_{\text{T}}\rangle/m_\chi$ at 50 cm$^2/$g (10 cm$^2/$g). The light blue shaded region highlights the bracketing region for  50 cm$^2/$g. Similarly, the solid and dashed brown lines bracket the ellipticity constraint, with the light brown shaded region highlighting the bracketed region.
  As the DM mass increases, Sommerfeld-enhanced indirect detection signals become increasingly effective at probing the parameter space. This is unsurprising, as the bulk of the high-mass parameter space is in the reannihilation regime, where the relatively large values of $\alpha_D$ and $\langle\sigma v\rangle$ accordingly yield interesting indirect detection signals.  The remaining unexcluded territory is predominantly in the more weakly coupled leak-in regime, where indirect detection signals are much fainter.  

For $m_\chi\gtrsim 100$ GeV,  we obtain viable parameter space realizing LIDM with DM self-interactions $\langle \sigma_{\text{T}}\rangle/m_\chi \sim$ few cm$^2/$g in dwarf systems, i.e., in the range of interest for addressing small-scale puzzles in galaxy formation, that are not obviously in tension with ellipticity constraints.

\section{Summary and conclusions}
\label{sec:conclusions}

In this paper we have examined in detail the properties of {\em leak-in dark matter}: dark matter that freezes out of a hidden sector evolving in a non-adiabatic leak-in phase.   The quasi-static equilibrium leak-in phase, in which the energy density of the hidden sector redshifts like matter, is a generic behavior that emerges when a cold hidden sector is dominantly populated through a dimension-four interaction with the hotter SM.   We provide analytic methods for consistently treating the out-of-equilibrium evolution of the hidden sector temperature 
in the presence of a known collision term. 

We present a detailed study of DM freezing out of a leak-in radiation bath and the resulting observational consequences.    The renormalizable nature of the interaction feeding the hidden sector radiation bath ensures that the cosmological evolution of the hidden sector is minimally sensitive to details of the unknown physics of reheating in our universe. This class of DM models are thus sharply predictive, and have a bounded parameter space. The strength of the interaction cannot be too large, in which case the interaction will reach equilibrium, or too small, in  which case the dark sector will never reach a high enough internal temperature to produce the observed DM relic abundance.  Meanwhile, the DM mass is bounded from above by the requirement of perturbativity, and from below by  a (model-dependent) combination of terrestrial, astrophysical, and cosmological constraints.  In an out-of-equilibrium hidden sector, the DM relic abundance is determined by an interplay of freezeout and freezein processes, resulting in a rich solution space. 

To establish some concrete constraints on and predictions from LIDM, we specialize to a particular model, where the dark sector consists of fermionic DM together with a dark vector boson that couples to the SM via the $B-L$ current.   Despite the smallness of the portal coupling $\epsilon$, there are many experimental probes of this $B-L$ LIDM model. 
While the DM annihilation cross-section is suppressed compared to standard WIMP scenarios thanks to the relative coldness of the hidden sector, indirect detection signals do not depend directly on the small portal coupling $\epsilon$, and provide excellent sensitivity to large regions of the parameter space.  
In particular, this model can realize very large DM masses ($m_\chi \sim$ 10s -- 100s of TeV) with striking cosmic ray signals of DM annihilation, detectable due to sizable Sommerfeld enhancements in the late universe from the relatively large dark coupling constant.  
Additionally, the enhanced cross-sections obtained from light mediator exchange enable direct detection experiments to probe the cosmic history, and not just the particle content, of thermal dark sectors.   In fact, XENON1T now provides the leading constraints on the very weakly coupled LIDM regime when $m_\zd<2 m_e$ and indirect detection signals are suppressed. 

Portions of the LIDM parameter space can realize very large DM self-interaction cross-sections.  The combination of (i) stringent constraints on low-mass $B-L$ gauge bosons, (ii) enormous DM self-interaction cross-sections, and (iii) the requirement of internal thermalization eliminates all parameter space where the $B-L$ boson lies below the constraints from stellar cooling, $\sim 100$ keV. Astrophysical tests of DM self-interactions could potentially provide a unique observational handle on the low-mass regions of LIDM parameter space, where neither direct nor indirect  detection
are sensitive.  Viable parameter space at high masses, $m_\chi \sim 10$ -- 100 GeV, 
can have DM 
self-interaction cross-sections
that fall 
in the astrophysically interesting range $\langle \sigma_{\text{T}}\rangle/m_\chi \sim$ few cm$^2/$g 
compatible with small-scale structure anomalies 
in dwarf systems.

 Leak-in dark matter represents a simple, generic, and sharply predictive class of models for the origin of dark matter in our universe.  For that reason, exploring the signature space of both this $B-L$ model and other realizations of LIDM, coming from other choices of leading interactions between the SM and the dark sector, is an important aspect of broadening the search for DM.

\bigskip

\noindent {\em Note added:}  While this work was nearing completion, the works Refs.~\cite{Hambye:2019dwd,Heeba:2019jho,Mohapatra:2019ysk} appeared, containing related but not identical material.

\bigskip

\noindent {\em Acknowledgements:} We gratefully thank J.~H.~Chang, J.~Cornell, G.~Holder, M.~Kaplinghat, C.~Kilic, and S.~Knapen  for useful conversations. The work of CG and JS is supported in part by DOE
Early Career grant DE-SC0017840.  JAE acknowledges support by DOE grant DE-SC0011784.   JAE and JS thank the Aspen Center for Physics under NSF grant PHY-1607611 for hospitality during the completion of this work.

\appendix

\section{Attaining internal thermalization}
\label{sec:inttherm}

In order for the dynamics described here to be an accurate description
of the hidden sector, the dark radiation bath must have sufficiently
rapid self-interactions to attain internal thermal equilibrium.  This
criterion depends on the properties of the dark radiation bath itself,
and is therefore necessarily somewhat model-dependent.  In this
subsection we will present an approximate criterion for internal
thermalization of the minimal $B-L$ vector portal hidden sector.

For the hidden sector to attain internal thermal equilibrium,
processes that change the numbers of individual dark species must be
efficient on cosmological timescales.  At leading order, such a
process is provided by the elastic scattering $\zd\zd\to
\chi\bar\chi$.  Given a number density $n_\zd$ of ``hard'',
pre-thermalized dark vectors, it is straightforward to estimate the
rate $\Gamma_{el}$ for this process. The number density of dark
photons in the absence of subsequent scattering within the hidden
sector can be obtained by solving the Boltzmann equation
\begin{equation}
\label{eq:nboltz}
\dot n_\zd + 3 H n_\zd = C(T)
\end{equation} 
under the simplifying assumptions that $H\propto T^2$ depends
only on the SM temperature, backward contributions to the collision
term can be neglected, and the SM temperature simply redshifts as
$T\propto 1/a$.  The collision term can be estimated as 
\beq
C(T)\approx
n_{g} (\sum_q n_q Q_q^2) \vev{\sigma_{qg\to qZ_D}  v} \sim \frac{
\epsilon^2 \alpha_s}{30} T^4,
\eeq 
where $\vev{\sigma_{qg\to qZ_D}  v}\sim \epsilon^2 \alpha_s/(24T^2)$ is the spin- and color- averaged cross-section.
Solving Eq.~\ref{eq:nboltz} yields
\begin{equation}
n_\zd \approx  \frac{ \epsilon^2 \alpha_s }{600} M_{Pl} T ^ 2.
\end{equation}
Comparing this result for $n_\zd$ to the analogous estimate for the
energy density injected into the HS (see Sec.~\ref{sec:analytics}), we
can see that (as expected) the typical energy carried by one of these
hard dark vectors is $\sim T$. The corresponding rate for initial
production of DM particles from the primordial dark vector population
is then
\begin{equation}
\Gamma_{el} \approx n_\zd  \times \frac{\pi\alpha_D^2}{T^2}\approx \frac{ \epsilon^2 \alpha_s\alpha_D^2}{20} M_{Pl}.
\end{equation} 
It is worth observing that the essential parametrics of this elastic
rate hold for any elastic $2\to 2$ process occurring among the initial
hard population of particles in the dark sector.  Given $\Gamma_{el}$, we
can quickly estimate whether elastic scattering is sufficiently rapid
to thermalize the hidden sector by requiring that $\Gamma_{el} > H$ at
some temperature $T> m_\chi$.  This estimate indicates that elastic
scattering suffices to thermalize much but not all of the
leak-in parameter space.

However, inelastic scattering, $\chi X \to \chi X + \zd$,
is more effective than elastic scattering at thermalizing the hidden
sector over much of the parameter space of interest. The importance of inelastic scattering in
thermalizing a sector containing gauge interactions is well-known 
\cite{Baier:1996vi,Peigne:2008wu,Mukaida:2015ria,Garny:2018grs}.   
While the inelastic scattering process is higher-order in $\alpha_D$,
it can be sufficiently enhanced by the region of low momentum
transfer to more than compensate for the additional $\alpha_D$ suppression.
Our estimate of thermalization through this inelastic process will be
parametric, and largely follows the related treatment in \cite{Garny:2018grs}.  

The inelastic scattering rate is approximately given by 
\begin{equation}
\label{eq:gammainel0}
\Gamma_{inel}  = n_\chi (\sigma v)_{inel} \sim n_\chi \,\frac{\pi\alpha_D^3}{\mu^2} ,
\end{equation}
where $\mu$ is the effective IR scale that regulates the $t$-channel
$\zd$ propagator, $n_\chi$ indicates the number density of hard DM
particles produced directly from the SM, and we have temporarily
neglected the possible complications that arise when the timescale for
emitting a soft vector boson in the final state becomes longer than
the timescale between hard scatterings,
i.e., 
the Landau-Pomeranchuk-Migdal (LPM) effect \cite{Landau:1953gr,Migdal:1956tc}.
The number density of hard $\chi$ particles, in the absence of subsequent
scattering within the hidden sector, can be obtained analogously to
the estimate for $n_\zd$ above.  We can estimate $\sv \approx (\sum_f
g_f (Q_{B-L}^f)^2 \, \pi\alpha_D \epsilon^2 /T^2$ for $f\bar f\to
\chi\bar\chi$, giving the collision term
\beq
C(T) \approx \left(\frac{3}{4}\frac{\zeta(3)}{\pi^2}\right)^2  \left(\sum_f g_f (Q_{B-L}^f)^2\right) \pi\alpha_D \epsilon^2 T^4  \sim \alpha_D\epsilon^2 T^4.
\eeq
Then, solving the Boltzmann equation for $n_\chi$ yields 
\begin{equation} 
\label{eq:nchi}
n_\chi (T) \sim \frac{\alpha_D\epsilon^2}{10} M_P T^2.
\end{equation}
%
%

\begin{figure}
\begin{center}
\includegraphics[width=0.8\textwidth]{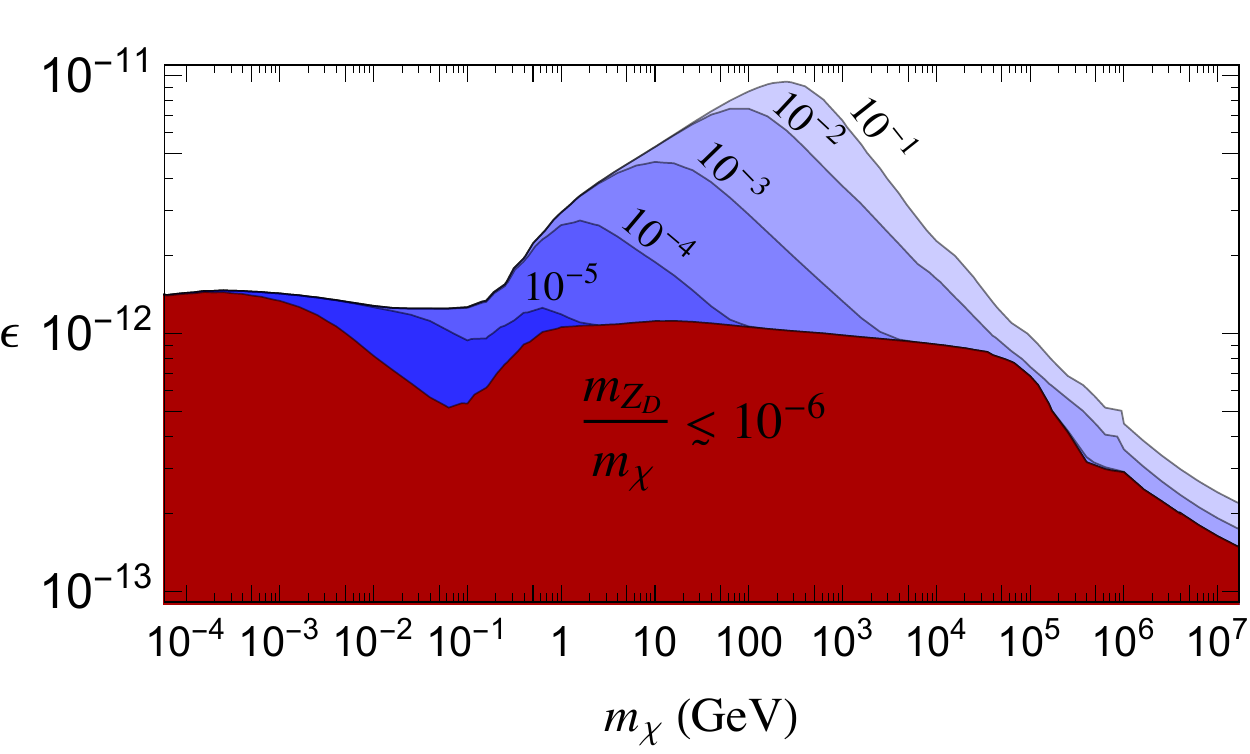}
\caption{The regions where our $B-L$ model does not internally thermalizes for different contours of $m_{Z_D}/m_\chi$.   As $m_{Z_D}$ gets closer to $m_\chi$, the inelastic scattering processes become less efficient.  For $m_{Z_D}/m_\chi\lesssim 10^{-6}$, the inelastic processes are maximally efficient and do not benefit from smaller mass ratios.  At low dark matter masses, the elastic processes dominate the thermalization, sculpting the region in the upper left part of the curve. 
}
\label{fig:therm}
\end{center}
\end{figure}

There are three possibilities for the effective IR scale $\mu$
that cuts off the momentum transfer in Eq.~\ref{eq:gammainel0}.  First
is simply  the (vacuum) dark vector mass itself, $m_\zd$. Second is $H$, 
reflecting that the horizon is the largest range of physical
interest for the dark interaction.  Finally, in the medium, the dark vector
propagator receives corrections from its interactions with the
plasma. The screening scale in the non-equilibrium dark plasma can be
estimated as \cite{Arnold:2002zm}
\begin{equation} 
\label{eq:IRscreening}
\mu_{sc}^2\approx \alpha_D \int \frac{d^3 p}{(2\pi)^3}\, \frac{f_\chi(p)}{p} \sim \frac{\alpha_D n_\chi}{T}, 
\end{equation}
in terms of the hard DM population $n_\chi$.
For the $B-L$ dark vector, we should in principle also
consider the contribution to its effective mass from interactions with
the SM plasma, $m_{SM,T}\sim \epsilon T$. 
Over our parameter range of interest, we
find that both Hubble and the SM contribution to the dark vector's effective mass 
are always negligible in comparison with $\mu_{sc}$ and $m_\zd$.
These possible screening scales have varying dependence on $T$,
$\epsilon$, and $\alpha_D$; at any given temperature, the {\em
  largest} is the one that is physically relevant.

Now, when the timescale for emitting a soft dark vector is larger than
the typical timescale between $2\to 2$ collisions, the inelastic $2\to 3$ scattering can no longer be discussed in isolation.
The result of multiple $2\to 2$ scatterings occurring during the so-called ``formation time'' governing the $1\to 2$ splitting is
known as the LPM
effect, and can be formally understood in an effective Boltzmann treatment
by defining an effective splitting function that resums specific
contributions to the amplitude from successive scatterings 
\cite{Arnold:2002zm,Kurkela:2011ti}.  Destructive interference 
among these contributions results in a suppression
of the brehmsstrahlung rate.  Thus we need to correct the
estimate of the inelastic rate for $2\to 3$ scattering in
Eq.~\ref{eq:gammainel0} with a factor $f_{LPM}\leq 1$ to account for
this suppression,
\begin{equation}
\label{eq:gammainel}
\Gamma_{inel} \sim n_\chi \,\frac{\pi\alpha_D^3}{\mu_{IR}^2} \times f_{LPM}.
\end{equation}
We use the estimate of \cite{Garny:2018grs} (see also
\cite{Mukaida:2015ria}) for $f_{LPM}$ in the Abelian plasma:
\begin{equation}
f_{LPM} \sim \mathrm{min} \left[1, \frac{\alpha_D \sqrt{n_\chi/T}}{n_\chi \,\pi\alpha_D^2/\mu_{IR}^2} \right]
\end{equation} 
provided $m_\zd\ll \alpha_D \sqrt{\left(n_\chi T \right)}$.  For
$m_\zd> \alpha_D \sqrt{\left(n_\chi T \right)}$, the LPM suppression
is not operative, so $f_{LPM} = 1$.  When the LPM effect is
operable, i.e., $f_{LPM}<1$, the net inelastic rate is given by $\Gamma_{inel} = \alpha_D^2
\sqrt{n_\chi/T}$.  If $m_{Z_D}$ is small, then the LPM effect is active everywhere in the parameter space of interest.   It is worth noting that this estimate for $f_{LPM}$ assumes an adiabatic evolution of $n_\chi$ in estimating the evolution of the formation timescale.  This is an {\em
  underestimate} of the non-adiabatic population of hard $n_\chi$, and
therefore an underestimate of $\Gamma_{inel}$.  While this treatment
could in principle be improved, it is a conservative choice, and
further refinement is beyond the scope of this paper.

In Fig.~\ref{fig:therm},  we show where the internal thermalization conditions are not satisfied, i.e.~when $\Gamma_{el} +\Gamma_{inel} < H$ at the freezeout temperature.  In practice, $2\to2$  processes are more important at lighter DM masses, while for higher DM masses the $2\to3$ process are more important. 
 In Figs.~\ref{fig:montanaroadmap} \& \ref{fig:alphaDsigma}, we show the $m_{Z_D}/m_\chi\lesssim 10^{-6}$ contours, while for Figs.~\ref{fig:mainparasp1} \& \ref{fig:mainparasp2}, we display the proper mass ratio dependent internal thermalization curve.

\section{Alternative UV models for hierarchical $B-L$ charges}
\label{sec:model}

Throughout this paper, we introduced an extremely large $B-L$ charge for DM to create a disparity in the $B-L$ vector boson's coupling to SM fields compared to dark matter.   This model has the advantage of having clear predictions and no UV sensitivity; however, the large DM charge invites model-building questions.  In this appendix, we present two simple models that provide an explanation for the hierarchical couplings of the dark $U(1)$ gauge boson to DM and the SM $B-L$ current, and briefly sketch the impact of the added states on the DM signatures.   Both models involve a $U(1)_{B-L}\times U(1)_{D}$ symmetry, with the first introducing kinetic mixing and the second introducing a Higgs state.
In both models, if the reheat temperature is too high, there is a danger that the $B-L$ vector could thermalize the SM and hidden sectors.

\subsection{Kinetic mixing with a heavy $B-L$ gauge boson}
\label{sec:KM}

This model has three $U(1)$ factors in the UV: a dark $U(1)$ gauge boson, a separate $U(1)_{B-L}$, and SM  hypercharge.   In the gauge basis, the Lagrangian describing the interactions of the dark $U(1)$, $\hat Z_{D}^{\mu}$, and the $U(1)_{B-L}$ boson, $\hat X^\mu$,  reads
\barray
\mathcal{L}&=& -\frac{1}{4} \hat Z_{D\mu\nu}  \hat Z_{D}^{\mu\nu}+ \frac{\hat\epsilon}{2} \hat Z_{D\mu\nu} \hat X^{\mu\nu} - \frac{1}{4} \hat X_{\mu\nu} \hat X^{\mu\nu} \\
&& + g_{B-L} \hat X_\mu J^\mu_{B-L} + g_D \hat Z_{D\mu} \bar \chi \gamma^\mu \chi ,
\earray
where $J^{\mu}_{B-L}$ is the SM $B-L$ current, and $\chi$ is the dark matter. In other words, we start with a model where, in the gauge basis, the dark gauge boson talks only to dark matter, and will inherit its couplings to the SM $B-L$ current through kinetic mixing with a new $B-L$ gauge boson.  We assume that this $B-L$ gauge boson gets a large mass through spontaneous symmetry breaking, $m_{X,0}$ (the origin of this mass term, Higgs or St\"uckelberg, is unimportant).

Making the customary field redefinition
\beq
\left(\begin {array}{c} \bar Z_D \\ \bar X \end{array}\right) = 
       \left(\begin {array}{cc} \sqrt{1-\hat\epsilon ^2} & 0 \\ -\hat\epsilon & 1 \end{array}\right)
      \left(\begin {array}{c} \hat Z_D \\ \hat X \end{array}\right),
\eeq
and redefining $g_D = \hat g_D/\sqrt{1-\hat\epsilon^2}$, yields diagonal kinetic
terms for the gauge bosons, and couplings to matter of the form
\beq
D_\mu = \partial_\mu + i g_D Q_D \bar Z_{D\mu}+ ig_{B-L} Q_{B-L} (\eta \bar Z_{D\mu} +\bar X_\mu). 
\eeq
Here we have defined
\beq
\eta =\frac{\hat\epsilon} {\sqrt{1-\hat\epsilon ^ 2}} .
\eeq
Given masses $m_{{Z_D},0}^2$  and $m_{X,0}^2$ for $\hat Z_D$ and $\hat X$, the resulting mass-squared matrix for $\bar Z_D$ and $\bar X$ is
\beq
\mathcal{M}^2_V = m_{X,0}^2 \left(\begin {array}{cc}   1 & \eta  \\
                          \eta & \eta^2 + \delta ^ 2 \end{array}\right)
\eeq
where $\delta^2 \equiv m_{{Z_D},0}^2/m_{X,0}^2$.  We will be interested in $\delta^2 \ll 1$.  This mixing matrix is diagonalized by
\beq
\label{eq:tan2a}
\tan 2\alpha =\frac{-2\eta}{ 1-\eta ^ 2-\delta ^ 2}.
\eeq
Expressing $\delta^2$ in terms of the eigenmass $m_{Z_D}$, we have 
\beq
\delta^2 = \frac{m_{{Z_D}}^2}{m_{X,0}^2}  \left(1+\frac{\eta^2}{1-m^2_{Z_D}/m^ 2_{X, 0}}\right)
   \approx   \frac{m_{{Z_D}}^2}{m_{X}^2} 
\eeq
where in the last step we expanded to leading order in $\hat\epsilon$ (assuming $m^2_D\ll m^2_X$).
Thus the mixing angle can be written as
\beq
\sin\alpha = \frac{\eta}{\sqrt{(1-\delta^2)^2+\eta^2}}, \phantom{space} \cos\alpha = \frac{1-\delta^2}{\sqrt{(1-\delta^2)^2+\eta^2}},
\eeq
giving the two eigenstate couplings to matter (to leading order in $\hat \epsilon$)
\barray
D_\mu &\supset&  i g_{B-L} Q_{B-L} \lp (\eta\cos \alpha - \sin\alpha) Z_{D,\mu} +\cos \alpha X_\mu \rp+ i g_{D} Q_D (\cos \alpha Z_{D,\mu} + \sin\alpha X_\mu) \nonumber \\
&\supset& ig_ {B-L} Q_{B-L}  \left( \frac{-\eta \delta^2}{\sqrt{(1-\delta^2)^2+\eta^2
}}Z_{D,\mu} + \frac{1-\delta^2}{\sqrt{(1-\delta^2)^2+\eta^2}} X_\mu \right) \\
&&+  i g_{D} Q_D\lp \frac{1-\delta^2}{\sqrt{(1-\delta^2)^2+\eta^2}} Z_{D,\mu} +  \frac{\eta}{\sqrt{(1-\delta^2)^2+\eta^2}}X_\mu \rp    
\earray
For $\delta, \hat\epsilon \ll 1$, the effective $Z_D$ coupling to the SM $B-L$ current is then the product of the underlying $g_{B-L}$ and two independent small parameters,
\beq
\epsilon \equiv g_{B-L}  \eta \delta^2 =g_{B-L}  \hat \epsilon \frac{m_D^2}{m_{X}^2}.
\eeq
which is the small portal coupling $\epsilon$ used throughout this work. 

Importantly in this model, the heavy $B-L$ vector couples dark matter to the SM particles at the same order as the lighter dark vector, which results in a cancellation of the leading amplitude for direct detection processes:
\barray
\hspace{-8mm}\abs{\mathcal M^{NR}} &=&  A F(E_R)\lp \frac{g_{B-L} g_D \eta \delta^2 }{2 m_N E_R + m_{Z_D}^2} -\frac{g_{B-L} g_D \eta}{2 m_N E_R + m_{X}^2} \rp \sim \epsilon g_D A F(E_R)    \frac{2 m_N E_R}{m_{Z_D}^4},
\earray
which is suppressed by $2 m_N E_R / m_{Z_D}^2$ relative to Eq.~\ref{eq:DDamp}.

\subsection{Dark mixed Higgs}
\label{sec:darkHiggs}

As before, this model has three $U(1)$ factors in the UV: a $U(1)_D$ gauge boson, a separate $U(1)_{B-L}$, and SM  hypercharge.  Additionally, we introduce a scalar field $\phi$ that has charges $\{Q_{D,\phi},Q_{B-L,\phi}\}$ under the $U(1)_D$ and $U(1)_{B-L}$ symmetries.    The terms in our Lagrangian important for this discussion are
\beq
\mathcal{L}= D_\mu \phi^* D^\mu \phi + \frac 12 m_{X,0}^2   \hat X_\mu  \hat X^\mu  + g_{B-L} \hat X_\mu J^\mu_{B-L} + \hat g_D \hat Z_{D\mu} \bar \chi \gamma^\mu \chi + V(\phi).
\eeq
where kinetic mixing is assumed to be absent. 
The mass for the vector $\hat X^\mu$ could arise from a St\"uckelberg or Higgs mechanism, but this origin is unimportant.  The gauge bosons couple to matter, notably $\phi$, through covariant derivatives of the form
\beq
D_\mu = \partial_\mu + i g_D Q_D \hat Z_{D\mu}+ ig_{B-L} Q_{B-L} \hat X_\mu. 
\eeq
In standard fashion, $V(\phi)$ results in a VEV for $\phi$, $\vev{\phi}=w$, so that our low-energy mass matrix has the form
\beq
\mathcal{M}^2_V = m_{X,0}^2 \left(\begin {array}{cc}   1 +\kappa^2 & \delta \kappa  \\
                        \delta \kappa &   \delta ^ 2 \end{array}\right)
\eeq
where $\kappa=g_{B-L}Q_{B-L,\phi} w/m_{X,0}$ and $\delta =g_{D}Q_{D,\phi} w/m_{X,0}$, and $\kappa,\delta\ll 1$.  Diagonalizing this matrix gives masses that are simply $m_X^2 \approx m_{X,0}^2(1 +\kappa^2)$ and  $m_{Z_D}^2 = g_{D}^2Q_{D}^2 w^2 + \order{w^4/m_{X,0}^2}$, and a mixing angle, $\sin\theta \sim \delta \kappa$.  After this the two eigenstates $X^\mu$ and $Z_D^\mu$ couple to matter as
\barray
 D_\mu &=& \partial_\mu + i g_D Q_D (Z_{D\mu} + \delta \kappa X_\mu) + ig_{B-L} Q_{B-L} \lp -\delta \kappa Z_{D\mu} + X_\mu\rp \\
 &\equiv& \partial_\mu + i g_D Q_D \lp Z_{D\mu} - \frac{\epsilon}{g_{B-L}} X_\mu \rp + i Q_{B-L} \lp \epsilon Z_{D\mu} +g_{B-L} X_\mu\rp
\earray
where we have defined 
\beq
\epsilon = -\lp g_{B-L}^2 Q_{B-L,\phi}\rp \lp g_D Q_{D,\phi}\rp \frac{w^2}{m_{X,0}^2}.
\eeq

Unlike the previous model, the heavy $B-L$ vector contributions to dark matter - SM interactions are unimportant in the IR.  In principle, the remaining scalar degree of freedom from $\phi$ could affect the model in a few ways.  It could be in the plasma, which could affect both $\tilde g_*$ and rates relevant for internal thermalization.  One way to reduce phenomenological consequences from $\phi$ would be to introduce a fairly small $Q_{D,\phi}$, which can allow for a very large separation between $m_\phi$ and $m_{Z_D}$.  With $m_\phi \gg m_\chi$, $\phi$ is effectively removed from the low-energy theory.

\section{Collision term}
\label{app:CE}

This Appendix collects details concerning the calculation of the energy transfer collision term $\mathcal{C}_E$ governing the temperature evolution of the hidden sector.

\subsection{Away from the equilibration floor}
\label{sec:derivation}

The hidden sector temperature $\tilde T$ can be numerically determined as a function of the SM temperature $T$ by the following procedure. For an internally thermalized hidden sector, the energy density stored there defines its temperature
\beq
\tilde \rho = \frac{\pi^2}{30} \tilde g_{*}(\tilde T) \tilde T^4.
\eeq
Differentiating this expression with respect to time gives
\beq
\dot{\tilde \rho} = \frac{\pi^2}{30} \tilde g_{*}(\tilde T) \tilde T^4 \lp \frac{4}{\tilde T}\frac{d\tilde T}{dt} + \frac{d\ln \tilde g_{*}(\tilde T)}{d\tilde T}\frac{d\tilde T}{dt}  \rp
\eeq
where the second term in parentheses is typically negligible, especially for minimal hidden sectors.  The first term can be simplified by using the relation
\beq
\frac{d\tilde T}{dt}= \frac{d\tilde T}{dT}\frac{dT}{da}\frac{da}{dt} =\frac{d\tilde T}{dT}\lp -T H \rp,
\eeq
which holds provided that (i) the SM  dominates the entropy in the universe and (ii) $g_{*S}$ is slowly varying, so that $T^3a^3 = $ const holds to a good approximation (near the QCD phase transition, this assumption will not be good).

A particularly useful variable is $\xi = \tilde T/T$, the ratio of hidden sector to SM temperatures.    Noting that $d\tilde T/dT=T\, d\xi/dT+\xi$, we can express Eq.~\ref{eq:b2} as
\beq
\frac{d\xi}{dT} = \frac{30 C_E(T,\tilde T)}{4 \pi^2 H(T) \tilde g_{*}(\xi T) \xi^3 T^5}.
\label{eq:xieq}
\eeq
Assuming again that $\tilde g_{*}$ is constant in the region of interest and the hidden sector is sufficiently cold so that the transfer of energy out of the hidden sector is negligible, $C_E(T,\tilde T)\sim C_E^f(T)$, we can solve Eq.~\ref{eq:xieq} to obtain
\beq
\xi(T) = \lp \int_{T_i}^{T} d\bar T \frac{30 C_E^f(\bar T)}{\pi^2 \tilde g_{*} H(\bar T) \bar T^5} \rp^{\frac 14} \propto \epsilon^{\frac 12}T^{-\frac 14}.
\eeq
Here we have used that $\xi(T_i)\ll \xi(T)$ which is always true in this 
model of interest for a sufficiently high value of $T_i$.

\subsection{Near the equilibration floor}
\label{sec:ceb}

Near the equilibration floor, the collision term in Eq.~\ref{eq:xieq} can be expanded in terms of a parameter $\delta = 1 -\xi$ that goes to 0 when the two sectors are equilibrated,
\beq
C_E(T,\delta) = C_E^f(T) - \sum_{n=0} C_{E,n}^b(T) \delta^n = - \sum_{n=1} C_{E,n}^b(T)\delta^n,
\eeq
where we have used that $C_{E,0}^b(T)=C_E^f(T)$.
Given the functions $C_{E,n}^b(T)$, the resulting equation,
\beq
\frac{d\xi}{dT} = \frac{-30 \sum_{n=1}  C_{E,n}^b(T) (1-\xi)^i}{4 \pi^2 H(T) \tilde g_{*}(\xi T) \xi^3 T^5},
\label{eq:xieq2}
\eeq
can be straightforwardly numerically integrated near the equilibration floor where the backward collision term  becomes important.

To derive the functions $C_{E,n}^b(T)$, we will (as throughout) use Maxwell-Boltzmann statistics.  The backward scattering piece of the collision term (\ref{eq:CollisionTerm}) can be written as
\beq
C_E^b(T,\tilde T) = \int d\Pi_i ( 2\pi) ^ 4\delta ^ 4 (\sum p_i) E_4 |\mathcal{M}(12\to 34)|^2 e^{-E_3/T}e^{-E_4/\tilde T}.
\label{eq:CEb}
\eeq
The collision term can be related to the cross-section $\sigma(s)$ for the given process using
\beq
\int d\Pi_1 d\Pi_2 ( 2\pi) ^ 4\delta ^ 4 (\sum p_i) |\mathcal{M}(12\to 34)|^2 = 4 g_1 g_2 s \lambda \sigma(s),
\eeq
where $\lambda = \frac 1{2s} \lp \lp s - m_3^2 - m_4^2 \rp^2 - 4 m_3^2 m_4^2\rp^{1/2}$ is the dimensionless two-body kinematic factor. We can thus write (\ref{eq:CEb}) as 
\beq
C_E^b(T,\tilde T) =  4 g_1 g_2 \int d\Pi_3 d\Pi_4 s \lambda(s)  \sigma(s) E_4   e^{-E_3/T}e^{-E_4/\tilde T}.
\label{eq:CEb2}
\eeq
Following Gondolo and Gelmini \cite{Gondolo:1990dk}, we  define $E_\pm = E_3 \pm E_4$, in terms of which the integral can be written,
\beq
C_E^b(T,\tilde T) =   \frac{g_1 g_2}{2 (2\pi)^4} \int ds dE_+ dE_-  s \lambda \sigma(s) \frac12 \lp E_+ - E_-\rp   e^{\frac{-\lp E_+ (T+\tilde T) + E_- (T-\tilde T) \rp}{2T\tilde T}},
\label{eq:CEb3}
\eeq
where the limits of integration are
\barray  
\nonumber
s &\geq& \mbox{Max}[m_1+m_2,m_3+m_4] \equiv x \\
E_+ &\geq& \sqrt s \\
E_+ R  -2 \lambda(s) \sqrt{E_+^2 -s} \leq E_- &\leq&  E_+ R +2 \lambda(s) \sqrt{E_+^2 -s}.
   \nonumber
\earray
where we have defined the dimensionless quantity $R \equiv \lp m_4^2-m_3^2 \rp/s$.  It is possible to integrate over $E_-$ in Eq.~\ref{eq:CEb3} analytically to yield
\barray  
C_E^b(T,\tilde T) &=&   \frac{g_1 g_2}{2 (2\pi)^4} \int ds dE_+  \frac{s \lambda(s)  \sigma(s) T\tilde T}{\lp T-\tilde T\rp^2} e^{\frac{-E_+ (T+\tilde T)}{2T\tilde T}} 
\label{eq:GGwdiffT}
\\\
&\times& \left\{ e^{\frac{(T-\tilde T)\lp E_+ R  -2 \lambda(s) \sqrt{E_+^2 -s}\rp}{2T\tilde T}} \lp E_+ (R-1) (T-\tilde T) -2\lp T\tilde T + \lambda (T-\tilde T) \sqrt{E_+^2 -s} \rp   \rp \right.
\nonumber
\\
&-& \left. e^{\frac{(T-\tilde T)\lp E_+ R  -2 \lambda(s) \sqrt{E_+^2 -s}\rp}{2T\tilde T}} \lp E_+ (R-1) (T-\tilde T) -2\lp T\tilde T - \lambda (T-\tilde T) \sqrt{E_+^2 -s} \rp   \rp \right\}.
   \nonumber
\earray
At this point, we  define $\delta = 1- \tilde T /T$ and expand the above expression in powers of $\delta$.  The terms in the resulting series can individually be integrated over $E_+$ analytically.  As $\kappa\equiv 1-R$ appears frequently in these expressions, it is convenient to replace $R$ with $\kappa$ below.  For any given scattering process, the backward collision term can thus be approximated using
\beq
C_{E,i}^b(T) = \int_{x}^{\infty} \frac{s \lambda^2 \sigma(s)}{(2\pi)^4} \lp A^{(1)}_i(s,T) K_1\lp\frac{\sqrt s}{T}\rp +A^{(2)}_i(s,T) K_2\lp\frac{\sqrt s}{T}\rp \rp ds
\eeq
where the functions $A^{(j)}_i(s,T)$ are given by
\barray
A^{(1)}_0(s,T) &=& 0 \nonumber \\
A^{(2)}_0(s,T) &=& s T \kappa \nonumber \\
A^{(1)}_1(s,T) &=& -\frac{1}2  s^{3/2} \kappa^2 \nonumber\\
A^{(2)}_1(s,T) &=& -\frac{1}2  s T (3 \kappa^2 +4 \lambda^2) \nonumber \\
A^{(1)}_2(s,T) &=& \frac{1}8  s^{3/2} \kappa \lp12 \lambda^2 + \kappa (3\kappa-4) \rp \nonumber\\
A^{(2)}_2(s,T) &=& \frac{s}{8T} \lp s \kappa^3 +4 T^2 \lp 3\kappa^2 (\kappa-1) +4(3\kappa -1) \lambda^2 \rp \rp \\
A^{(1)}_3(s,T) &=& -\frac{s^{3/2}}{48T^2}\lp s \kappa^4 + 3T^2(8\kappa^2-12\kappa^3 +5\kappa^4 +8\kappa \lambda^2(5\kappa -6) +16\lambda^4 ) \rp \nonumber\\
A^{(2)}_3(s,T) &=& - \frac{s^2 \kappa^2}{8T} (\kappa^2 -2\kappa +4\lambda^2) - \frac{s T}{4}\!\! \lp 6\kappa^2 -12 \kappa^3 +5 \kappa^4 +8(5\kappa^2-6\kappa+1) \lambda^2  +16\lambda^4 \rp \nonumber \\
A^{(1)}_4(s,T) &=&  \frac{s^{3/2}}{64}\lp \kappa^2(15\kappa^3-60\kappa^2+72\kappa-32)  +8\kappa\lambda^2(5\kappa -6)^2 +48 \lambda^4 (5\kappa -4)  \rp \nonumber\\
& & + \frac{s^{5/2}\kappa^3}{192T^2}\lp 3\kappa^2-12\kappa+ 20\lambda^2 \rp \nonumber \\
A^{(2)}_4(s,T) &=& \frac{s}{384T^3} \lp s^2 \kappa^5 +3sT^2\kappa(48 \kappa^2 (1-\kappa) +13\kappa^4 +\kappa\lambda^2 (120\kappa-192) + 80\lambda^4) \rp \nonumber \\
& & + \frac{sT}{16} \lp3 \kappa^2(5\kappa^3 -20\kappa^2 +24\kappa -8)+8\lambda^2 (25\kappa^3-60\kappa^2 +36\kappa -4)+48 \lambda^4(5\kappa-4)  \rp \nonumber
\earray
We provide the complete expansion terms up to fourth order in $\delta$ as this proved sufficient to match onto the full solution up through the range with a fairly large value of $\delta =2/3$ ($\xi\sim 1/3$) where the backwards collision term could be reliably ignored.


\bibliography{DarkSectors2}
\bibliographystyle{JHEP}


\end{document}